\documentclass[preprint,12pt]{elsarticle}
\journal{Computer Physics Communications}

\usepackage{amssymb}
\usepackage{hyperref}

\newcommand{\brm}{\mathrm{b}}
\newcommand{\drm}{\mathrm{d}}
\newcommand{\erm}{\mathrm{e}}
\newcommand{\frm}{\mathrm{f}}
\newcommand{\grm}{\mathrm{g}}
\newcommand{\prm}{\mathrm{p}}
\newcommand{\qrm}{\mathrm{q}}
\newcommand{\trm}{\mathrm{t}}
\newcommand{\Brm}{\mathrm{B}}
\newcommand{\Drm}{\mathrm{D}}
\newcommand{\Jrm}{\mathrm{J}}
\newcommand{\Hrm}{\mathrm{H}}
\newcommand{\Wrm}{\mathrm{W}}
\newcommand{\Zrm}{\mathrm{Z}}
\newcommand{\bbar}{\overline{\mathrm{b}}}
\newcommand{\fbar}{\overline{\mathrm{f}}}
\newcommand{\pbar}{\overline{\mathrm{p}}}
\newcommand{\qbar}{\overline{\mathrm{q}}}
\newcommand{\MSbar}{\overline{\mathrm{MS}}}
\newcommand{\pT}{p_{\perp}}
\newcommand{\pTo}{p_{\perp 0}}
\newcommand{\cindent}{\hspace*{10mm}~}
\newcommand{\alphas}{\alpha_{\mathrm{s}}}

\begin{document}
\begin{frontmatter}
 
\title{
\begin{flushright}
{\normalsize
LU TP 14-36\\
MCNET-14-22\\
CERN-PH-TH-2014-190\\
FERMILAB-PUB-14-316-CD \\
DESY 14-178\\
SLAC-PUB-16122\\
October 2014}
\end{flushright}
An Introduction to PYTHIA 8.2}

\author[a]{Torbj\"orn~Sj\"ostrand\corref{corauthor}}
\author[b]{Stefan~Ask\fnref{nowatone}}
\author[a]{Jesper~R.~Christiansen}
\author[a]{Richard~Corke\fnref{nowattwo}}
\author[c]{Nishita~Desai}
\author[d]{Philip~Ilten}
\author[e]{Stephen~Mrenna}
\author[f,g]{Stefan~Prestel}
\author[a]{Christine~O.~Rasmussen}
\author[h,i]{Peter~Z.~Skands}

\cortext[corauthor]{Corresponding author; 
\textit{e-mail address:} \texttt{torbjorn@thep.lu.se}}
\fntext[nowatone]{Now at Winton Capital Management, Zurich, Switzerland}
\fntext[nowattwo]{Now at Nordea Bank, Copenhagen, Denmark}

\address[a]{Department of Astronomy and Theoretical Physics, Lund University,\\ 
S\"olvegatan 14A, SE-223 62 Lund, Sweden}
\address[b]{Department of Physics, University of Cambridge, Cambridge, UK}
\address[c]{Institut f\"ur Theoretische Physik, Universit\"at Heidelberg,\\
Philosophenweg 16, D-69120 Heidelberg, Germany}
\address[d]{Massachusetts Institute of Technology, Cambridge, MA 02139, USA}
\address[e]{Fermi National Accelerator Laboratory, Batavia, 
IL 60510, USA}
\address[f]{Theory Group, DESY, Notkestrasse 85, D-22607 Hamburg, 
Germany}
\address[g]{SLAC National Accelerator Laboratory, 
  Menlo Park, CA 94025, USA}
\address[h]{CERN/PH, CH--1211 Geneva 23, Switzerland}
\address[i]{School of Physics, Monash University, PO Box 27, 3800 Melbourne, 
Australia}

\begin{abstract}
The \textsc{Pythia} program is a standard tool for the generation 
of events in high-energy collisions, comprising a coherent set of physics
models for the evolution from a few-body hard process to a complex
multiparticle final state. It contains a library of hard processes, 
models for initial- and final-state parton showers, matching and merging
methods between hard processes and parton showers, multiparton 
interactions, beam remnants, string fragmentation and 
particle decays. It also has a set of utilities and several interfaces 
to external programs. \textsc{Pythia}~8.2 is the second main release after the 
complete rewrite from Fortran to C++, and now has reached such a
maturity that it offers a complete replacement for most applications, 
notably for LHC physics studies. The many new features should allow 
an improved description of data.

\end{abstract}


\end{frontmatter}

\clearpage


\noindent{\bf NEW VERSION PROGRAM SUMMARY}

\begin{small}
\noindent
{\em Manuscript Title: An Introduction to \textsc{Pythia} 8.2}  \\
{\em Authors: Torbj\"orn Sj\"ostrand, Stefan Ask, Jesper R. Christiansen,
Richard Corke, Nishita Desai, Philip Ilten, Stephen Mrenna, 
Stefan Prestel, Christine O. Rasmussen, Peter Z. Skands} \\
{\em Program Title:} \textsc{Pythia} 8.2                      \\
{\em Journal Reference:}                                      \\
{\em Catalogue identifier:}                                   \\
{\em Licensing provisions:} GPL version 2                     \\
{\em Programming language:} C++                               \\
{\em Computer:} commodity PCs, Macs                           \\
{\em Operating systems:} Linux, OS X; should also work on other systems \\
{\em RAM:} $\sim$10 megabytes                                    
{\em Classification:} 11.2  Phase Space and Event Simulation   \\
{\em Catalogue identifier of previous version:} ACTU\_v3\_0       \\
{\em Journal reference of previous version:} T.~Sj\"ostrand, S.~Mrenna 
and P.~Skands, Computer Physics Commun. {\bf 178} (2008) 852  \\
{\em Does the new version supersede the previous version?:} yes \\
{\em Nature of problem:} high-energy collisions between 
elementary particles normally give rise to complex final states,
with large multiplicities of hadrons, leptons, photons and neutrinos.
The relation between these final states and the underlying 
physics description is not a simple one, for two main reasons. 
Firstly, we do not even in principle have a complete understanding 
of the physics. Secondly, any analytical approach is made 
intractable by the large multiplicities.\\
{\em Solution method:} complete events are generated by Monte Carlo 
methods. The complexity is mastered by a subdivision of the full 
problem into a set of simpler separate tasks.
All main aspects of the events are simulated, such as
hard-process selection, initial- and final-state radiation, beam
remnants, fragmentation, decays, and so on. Therefore events should be
directly comparable with experimentally observable ones. The programs
can be used to extract physics from comparisons with existing
data, or to study physics at future experiments.\\
{\em Reasons for the new version:} improved and expanded physics models\\
{\em Summary of revisions:} hundreds of new features and bug fixes, 
allowing an improved modeling\\
{\em Restrictions:} depends on the problem studied\\
{\em Unusual features:} none\\
{\em Running time:} 10--1000 events per second, depending on
process studied   
\end{small}

\clearpage

\sloppy

\section{Introduction}

The \textsc{Pythia} program is a standard tool for the generation of
events in high-energy collisions between elementary particles, comprising a 
coherent set of physics models for the evolution from a few-body 
hard-scattering process to a complex multiparticle final state. 
Parts of the physics have been rigorously derived from theory, 
while other parts are based on phenomenological models, with parameters 
to be determined from data. Currently the largest user community 
can be found among the LHC experimentalists, 
but the program is also used for a multitude of other phenomenological or
experimental studies. Main tasks performed by the program
include the exploration of experimental 
consequences of theoretical models, the development of search 
strategies, the interpretation of experimental data, and the study 
of detector performance. Thereby it spans the whole lifetime of an
experiment, from early design concepts for the detector
to final presentation of data. 

The development of \textsc{Jetset} 
\cite{Sjostrand:1982fn,Sjostrand:1982am,Sjostrand:1985ys,Sjostrand:1986hx}
began in 1978 and many of its components were merged later with 
\textsc{Pythia} \cite{Bengtsson:1982jr,Bengtsson:1984yx,Bengtsson:1987kr,%
Sjostrand:1993yb,Sjostrand:2000wi,Sjostrand:2006za}. 
Thus the current \textsc{Pythia} generator is the product of more than 
35 years of development. 

{}At the onset, all code was written in Fortran~77. 
\textsc{Pythia}~8.100 \cite{Sjostrand:2007gs} was the first full release 
of a complete rewrite to C++. As such, some relevant features were 
still missing, and the code had not yet been tested and tuned 
to the same level of maturity as \textsc{Pythia}~6.4
\cite{Sjostrand:2006za}, the last version of the old Fortran generation. 
Since then missing features have been added, new features introduced, and
bugs found and fixed.  Experience has been building slowly in the 
experimental community and the LHC collaborations, in particular, 
are in the midst of or have made the full-scale transition from 
\textsc{Pythia}~6 to \textsc{Pythia}~8. 

The development of \textsc{Pythia}~8.1 after the original release 
has been a continuous process, and backwards
incompatibility has been introduced in only a few instances. 
We take the opportunity 
of the \textsc{Pythia}~8.200 release to introduce a further set of minor
incompatibilities, in particular to remove some outdated functionality.
Most user programs should work 
unchanged, or only require minimal adjustments. 

On the one hand, \textsc{Pythia} is intended to be self-contained, 
useful for any number of standalone physics studies. On the other,  
an ongoing trend is that \textsc{Pythia}~8 is interlinked with
other program packages. This is accomplished through 
a number of interfaces, with the Les Houches Accord (LHA) 
\cite{Boos:2001cv} and its associated Les Houches Event Files (LHEF) 
\cite{Alwall:2006yp,Butterworth:2014efa} being a prime example. In this
way, matrix-element (ME) based 
calculations from a number of different sources can be combined with
\textsc{Pythia} specialities such as initial-state radiation (ISR), 
final-state radiation (FSR), multiparton interactions (MPI), and 
string fragmentation.

The intention of this article is neither to provide a complete overview of 
the physics implemented, nor a complete user manual describing, for
example, the
complete set of runtime settings.  This was done for 
\textsc{Pythia}~6.4, and required 580 pages \cite{Sjostrand:2006za}. 
For \textsc{Pythia}~8 the user manual part is completely covered by a 
set of interlinked HTML (or alternatively PHP) pages that is distributed 
along with the program code. In addition there is a worksheet, 
intended for summer schools or self-study, that offers an introductory
tutorial, and example main programs to get started with various tasks. 

For the physics implementation, the story is more complex. Major parts of
the \textsc{Pythia}~6.4 writeup still are relevant, but there are also 
parts that have evolved further since. This is only briefly 
covered in the HTML manual. In the future we intend to link more PDF 
documents with detailed physics descriptions to it, but this will be 
a slow buildup process.   

The intention here is to provide information that explains the
evolution of the current program and makes the other resources,
such as the online manual, intelligible.
Section \ref{sec:physics} contains a concise summary of the physics of 
\textsc{Pythia}~8, with emphasis on limitations and on aspects that
are new since \textsc{Pythia}~8.100. A short overview of the program 
code follows in section \ref{sec:code}
that includes installation instructions, 
an outline of the main program elements, 
methods for user interaction with the code, the possibility
of interfaces with external 
libraries, and more. Section \ref{sec:outlook} rounds off the article
with an outlook to future developments.

\section{Physics Summary \label{sec:physics}}

As described in the introduction,
only a brief outline of the physics content will be provided here, 
with emphasis on those aspects that are new since \textsc{Pythia}~8.100 
and on \textsc{Pythia}'s area of application, since many 
user questions concern what \textsc{Pythia} \emph{cannot} do. 
Further details are available in the HTML manual, and in a number of
physics publications over the years, notably the \textsc{Pythia}~6.4
manual \cite{Sjostrand:2006za}.

\subsection{Limitations}

The physics models embodied in \textsc{Pythia} focus on
\emph{high-energy} particle collisions, defined as having
centre-of-mass (CM) energies greater than 10 GeV, corresponding 
to a proton--proton ($\prm\prm$) fixed-target beam energy of $\ge$ 50 GeV.  
This limitation is due to the approximation of a continuum of allowed 
final states being used in several places in \textsc{Pythia}, most
notably for hadron--hadron cross-section calculations, total and
differential, and as the basis for the string-fragmentation model. 
At energies below 10 GeV, we enter the hadronic resonance region, where
these approximations break down, and hence the results produced by
\textsc{Pythia} would not be reliable. The 10 GeV limit is picked as
a typical scale; for positron--electron ($\erm^+\erm^-$) annihilation 
it would be possible
to go somewhat lower, whereas for $\prm\prm$ collisions the models are
not particularly trustworthy near the lower limit.

At the opposite extreme, we are only aware of explicit tests of 
\textsc{Pythia}'s physics modeling up to CM energies of roughly 
100 TeV, corresponding to a $\prm\prm$ fixed-target beam energy 
$\le$ $10^{10}$ GeV; see e.g.~\cite{Skands:2013asa,Skands:2014pea}. 
Using \textsc{Pythia} to extrapolate  to even higher energies is not 
advised for novice users and should be accompanied by careful cross checks 
of the modeling and results. 

Currently the program only works either with \emph{hadron--hadron} or 
\emph{lepton--lepton} collisions. \emph{Hadron} here includes the
(anti)proton, 
(anti)neutron, pion and, as a special case, the Pomeron.
There is not yet any provision for lepton--hadron collisions 
or for incoming photon beams, though these could conceivably be added in 
future significant updates. Internal facilities to handle proton--nucleus 
or nucleus--nucleus collisions are not foreseen at all. For completeness, 
however, we note that a wide range of programs exist with
interfaces to some of the physics models in \textsc{Pythia}, 
in particular the string fragmentation routines,
for collision and decay processes.

The outgoing particles are produced in vacuum and the
simulation 
of the interaction of the produced particles with \emph{detector 
material is not included} in \textsc{Pythia}.
Interfaces to external detector-simulation codes can be written
directly by the user or 
accomplished via the HepMC \cite{Dobbs:2001ck} interface, 
as described in 
subsection \ref{subsubsec:hepmc}. 
Analysis of \textsc{Pythia} events can always be done 
at the parton or particle level.   Examples of such analyses
are provided with the code distribution.

\subsection{Hard processes}

A large number of processes are available internally, and even more through
interfaces to external programs. The input of external hard processes
via the LHA/LHEF standards actually is the main source of a rapidly
expanding set of processes that \textsc{Pythia} can handle. Nevertheless
there will always remain some need for internal processes, in part the 
standard ones required for basic physics studies, in part ones with 
special requirements, like involving long-lived coloured particles, 
new colour structures, or parton showers in new gauge groups.
Recent internal additions include several scenarios for Hidden Valley 
physics, further processes involving 
extra dimensions, more Supersymmetric (SUSY) 
processes, extended
handling of $R$-hadrons, and more 
charmonium/bottomonium states.

The implementation of hard processes focuses on $2 \rightarrow 1$ and 
$2 \rightarrow 2$ processes with some $2\rightarrow 3$ processes available.  
It may be possible however, to generate processes with higher final-state 
multiplicity if the particles arise from decays of resonances. As of version 
8.2, the following processes are available internally:

\begin{itemize}

\item {\bf QCD processes} include both soft- and hard-QCD processes.
  The hard-QCD processes include the standard $2\to 2$ ones available
  in \textsc{Pythia} 6.4, 
  with open charm and bottom production,
  as well as new $2\to 3$ processes that can
  be used, for example, for comparisons with parton showers.
  
\item {\bf Electroweak (EW) processes} include prompt photon production,
  single production of $\gamma^* /\Zrm$ and $\Wrm^\pm$ as well as pair
  production of weak bosons with full fermion correlations for $VV
  \rightarrow 4\frm$.  Photon collision processes of the type 
  $\gamma \gamma \rightarrow \frm \fbar$ are also available.
  
\item {\bf Onia} include production of any $^3S_1$, $^3P_J$ and
  $^3D_J$ states of charmonium or bottomonium via colour-singlet and
  colour-octet mechanisms.
  
\item {\bf Top} production, singly or in pairs.

\item {\bf Fourth generation} fermion production via strong or
  EW interactions.
  
\item {\bf Higgs processes} include the production of the SM Higgs boson
  as well as the multiple Higgs bosons of
  a generic two-Higgs-doublet model (2HDM), with the
  possibility of $CP$ violating decays.  It is also possible to
  modify the angular correlation of the Higgs decay $h \rightarrow
  VV \rightarrow 4\frm$ due to anomalous $hVV$ couplings. The internal
  implementation of SUSY also uses the 2HDM implementation
  for its Higgs sector.
  
\item {\bf SUSY processes} include the pair production of SUSY
  particles as well as resonant production of squarks via the $R$-parity
  violating UDD interaction. EW interferences have been taken into
  account where relevant and can be turned off for comparisons with
  \textsc{Pythia}~6.4. The implementation has been documented in
  \cite{Desai:2011su}. Both squarks and gluinos can be made to form
  long-lived $R$-hadrons, that subsequently decay. In between it is 
  possible to change the ordinary-flavour content of the $R$-hadrons, 
  by (user-implemented) interactions with the detector material 
  \cite{Fairbairn:2006gg}.
  
\item {\bf New gauge boson processes} include production of a $\Zrm'$
  (with full $\gamma^*/\Zrm/\Zrm'$ interference), a $\Wrm'^{\pm}$ and of a
  horizontally-coupling (between generations) gauge boson $\mathrm{R}^0$.
  
\item {\bf Left-right symmetric processes} include the production of the
  $SU(2)_R$ bosons $\Wrm^\pm_R$, $\Zrm^0_R$, and the doubly charged
  Higgs bosons $\Hrm_L^{++}$ and $\Hrm_R^{++}$.
  
\item { \bf Leptoquark production}, singly or in pairs, 
  with the assumption that the leptoquark always
  decays before fragmentation.
  
\item { \bf Compositeness processes} include the production of excited fermions
  and the presence of contact interactions in QCD or EW processes.
  The production of excited fermions can be via both gauge and contact
  interactions; however, only decays via gauge interactions are
  supported with angular correlation.
  
\item {\bf Hidden Valley processes} can be used to study visible
  consequences of radiation in a hidden sector. Showering is modified
  to include a third kind of radiation, fully interleaved with the QCD
  and QED radiation of the SM.  New particles include
  $SU(N)$-charged gauge bosons as well as partners of the SM
  fermions charged under $SU(N)$.  See
  \cite{Carloni:2010tw,Carloni:2011kk} for further details.
  
\item {\bf Extra-dimension processes} include the production of particles
  predicted by Randall-Sundrum models, TeV-sized and Large Extra
  Dimensions, and Unparticles. See \cite{ Ask:2008fh, Ask:2009pv,
    Ask:2011zs} for detailed descriptions.
    
\end{itemize}

The full list of available processes and parameters for BSM models along 
with references is available in the HTML manual distributed with the code.  
Furthermore, for the cases of one, two, or three hard partons/particles in 
the final state, the user can also use the {\sc Pythia} class structure to code 
matrix elements for required processes as yet unavailable internally, 
and even use \textsc{MadGraph}~5 \cite{Alwall:2011uj} to automatically generate
such code. This is discussed later in subsection \ref{sec:semi-internal}.

\subsection{Soft processes}

\textsc{Pythia} is intended to describe all components of the total
cross section in hadronic collisions, including elastic, diffractive
and non-diffractive topologies. Traditionally special emphasis is put 
on the latter class, which constitutes the major part of the total 
cross section. In recent years the modeling of diffraction has improved
to a comparable level, even if tuning of the related free parameters is 
lagging behind.

The total, elastic, and inelastic cross
sections are obtained from Regge fits to data. At the time of writing, the
default for $\prm\prm$ collisions is the 1992 Donnachie-Landshoff 
parametrisation~\cite{Donnachie:1992ny},  with one Pomeron and one
Reggeon term, 
\begin{equation}
\sigma^{\prm\prm}_\mathrm{TOT}(s) \ = \ 
   (21.70\, s^{0.0808}
 + 56.08\, s^{-0.4525})~\mathrm{mb},
\end{equation}
with the $\prm\prm$ CM energy squared, $s$, in units of
$\mathrm{GeV}^2$. For $\prm\pbar$ collisions, 
the coefficient of the second (Reggeon) term changes to
$98.39$; see~\cite{Donnachie:1992ny,Schuler:1993td,Sjostrand:2006za}
for other beam types. 

The elastic cross section is approximated by a
simple exponential falloff with momentum transfer, 
valid at small Mandelstam $t$, related to the total cross section via
the optical theorem,  
\begin{equation}
\frac{\drm\sigma^{\prm\prm}_\mathrm{EL}(s)}
     {\drm t} \ = \ \frac{(\sigma^{\prm\prm}_\mathrm{TOT})^2}{16\pi}
 \exp\left(B^{\prm\prm}_\mathrm{EL}(s)\, t\right)
\hspace*{0.5cm}\to\hspace*{0.5cm}~
\sigma^{\prm\prm}_\mathrm{EL}(s) \ =
\ \frac{(\sigma^{\prm\prm}_\mathrm{TOT})^2}{16\pi
  B^{\prm\prm}_\mathrm{EL}(s)} 
~,
\end{equation}
using $1~\mathrm{mb} = 1/(0.3894\, \mathrm{GeV}^2)$ 
to convert between $\mathrm{mb}$ and
$\mathrm{GeV}$ units, and $B_\mathrm{EL}^{\prm\prm} = 5 + 4s^{0.0808}$ the 
$\prm\prm$ elastic slope in $\mathrm{GeV}^{-2}$, defined using the same power
of $s$ as the Pomeron 
term in $\sigma_\mathrm{TOT}$, to maintain sensible
asymptotic behaviour at high energies. We emphasise that also the
electromagnetic Coulomb term, with  interference, can optionally 
be switched on for elastic scattering --- a feature so far unique to
\textsc{Pythia} among major generators. 

The inelastic cross section is a derived quantity:
\begin{equation}
\sigma_\mathrm{INEL}(s) = \sigma_\mathrm{TOT}(s) -
\sigma_\mathrm{EL}(s)~.\label{eq:sigINEL}
\end{equation}
The relative breakdown of the inelastic cross
section into single-diffractive (SD), double-diffractive (DD),
central-diffractive (CD), and non-diffractive (ND) components
is given by a choice between 5 different
parametrisations~\cite{Navin:2010kk,Ciesielski:2012mc}. 
The current default is the Schuler-Sj\"ostrand
one~\cite{Schuler:1993td,Schuler:1993wr}: 
\begin{eqnarray}
\frac{\drm\sigma^{\prm\prm\to X\prm}_{\mathrm{SD}}(s)}{\drm t\,\drm M_X^2} 
 & = &
 \frac{g_{3\mathbb{P}}}{16\pi}
 \frac{\beta_{\prm\mathbb{P}}^3}{M_X^2}\, F_\mathrm{SD}(M_X)\, 
\exp\left(B_{\mathrm{SD}}^{X\prm}\,t\right)~,
\\[2mm]
\frac{\drm\sigma_\mathrm{DD}^{\prm\prm}(s)}{\drm t\,\drm M_1^2\,\drm M_2^2} 
 & = & 
 \frac{g_{3\mathbb{P}}^2}{16\pi}
\frac{\beta_{\prm\mathbb{P}}^2}{M_1^2M_2^2}\,F_\mathrm{DD}(M_1,M_2)\,
\exp\left(B_{\mathrm{DD}}\,t\right)~,
\end{eqnarray}
with the diffractive masses ($M_X$, $M_1$, $M_2$),   
the Pomeron couplings ($g_{3\mathbb{P}}$, $\beta_{\prm\mathbb{P}}$),
the diffractive slopes ($B_\mathrm{SD}$, $B_\mathrm{DD}$), and
the low-mass resonance-region enhancement and high-mass kinematical-limit
suppression factors ($F_\mathrm{SD}$, $F_\mathrm{DD}$) summarised 
in~\cite{Navin:2010kk}.  

The central-diffractive component is a new
addition, not originally included in \cite{Navin:2010kk}. 
By default, it is parametrised according to a simple scaling
assumption, 
\begin{equation}
\sigma_\mathrm{CD}(s)
 =  \sigma_\mathrm{CD}(s_\mathrm{ref})
\left(\frac{\ln(0.06~s/s_0)}{\ln(0.06~s_\mathrm{ref}/s_0)}\right)^{3/2}~,
\end{equation}
with $\sigma_\mathrm{CD}(s_\mathrm{ref})$ the CD cross
section at a fixed reference CM energy chosen to be 
$\sqrt{s_{\mathrm{ref}}}=2\,\mathrm{TeV}$ by default and 
$\sqrt{s_0}=1\,\mathrm{GeV}$. The spectrum is 
distributed according to 
\begin{equation}
\frac{\drm\sigma^{\prm\prm}_{\mathrm{CD}}(s)}%
  {\drm t_1\,\drm t_2\,\drm\xi_1\,\drm\xi_2}  
  \propto  \frac{1}{\xi_1\xi_2}\,
\exp\left(B_\mathrm{SD}^{\prm X}\,t_1\right)\exp\left(B_\mathrm{SD}^{X\prm}
\,t_2\right)~,
\end{equation}
with $\xi_{1,2}$ being the fraction of the proton energy 
carried away by the Pomeron, related to the diffractive mass through 
$M_\mathrm{CD}=\sqrt{\xi_1\xi_2 s}$.

Depending on the selected diffractive parametrisation, the
non-diffractive cross section is evaluated by integrating the
diffractive components and subtracting them from $\sigma_\mathrm{INEL}$,
\begin{eqnarray}
\sigma^{\prm\prm}_\mathrm{ND}(s) & = & \sigma^{\prm\prm}_\mathrm{INEL}(s) 
- \int \left(\drm\sigma_{\mathrm{SD}}^{\prm\prm\to X\prm}(s) 
+ \drm\sigma_{\mathrm{SD}}^{\prm\prm\to \prm X}(s) 
+ \drm\sigma_\mathrm{DD}^{\prm\prm}(s) + \drm\sigma_\mathrm{CD}^{\prm\prm}(s)
\right)~.\label{eq:sigND}
\end{eqnarray}
Note, therefore, that the ND cross section is
only defined implicitly, via eqs.~(\ref{eq:sigINEL}) --
(\ref{eq:sigND}).  

We emphasise that recent precision measurements at high energies, in
particular by 
TOTEM~\cite{Antchev:2013iaa,Antchev:2013paa} and by
ALPHA~\cite{Aad:2014dca}, 
have highlighted that $\sigma_\mathrm{TOT}(s)$ and
$\sigma_\mathrm{EL}(s)$ actually grow a bit
faster at large $s$, while $\sigma_\mathrm{INEL}(s)$ remains in the
right ballpark. More recent
fits~\cite{Cudell:1996sh,Donnachie:2013xia} are consistent with
using a power $s^{0.096}$ for the Pomeron term. 
Updating the
total cross-section formulae in \textsc{Pythia}~8 is on the to-do list for a
future revision. 

Alternatively, it is also possible to set your own user-defined cross
sections (values only, not functional forms), see the HTML manual's
section on ``Total Cross Sections''.

Among the event classes, the non-diffractive one is the norm, in the 
context of which most aspects of event generators have been developed. 
It is therefore amply covered in subsequent sections. 

Single, double and central diffraction now are handled in the spirit of 
the Ingelman--Schlein model \cite{Ingelman:1984ns}, wherein a Pomeron 
is viewed as glueball-like hadronic state. The Pomeron flux defines the 
mass spectrum of diffractive systems, whereas the internal structure of 
this system is simulated in the spirit of a non-diffractive hadronic 
collision between a Pomeron and a proton \cite{Navin:2010kk}. 
Low-mass diffractive systems are still assumed to exhibit no 
perturbative effects and hence are represented as purely non-perturbative
hadronizing strings, respecting the quantum numbers of the diffractively
excited hadrons and with phenomenological parameters governing the choice
between two different possible string configurations. 
For diffractive systems with masses greater than about 10 GeV (a
user-modifiable smooth transition scale), ISR and FSR effects are
fully included, hence diffractive 
jets are showered, and the additional possibility of MPI within the
Pomeron--proton system allows for an underlying event to be generated
within the diffractive system.   

Exclusive diffractive processes, like $\prm \prm \to \prm \prm h$, with $h$ 
representing a single hadron, have \textit{not} been implemented 
and would in any case not profit from the full \textsc{Pythia} machinery.

\subsection{Parton distributions}

Currently, sixteen parton distribution function (PDF) sets for the 
proton come built-in.  In addition to the internal proton sets, a few
sets are also available for the pion, the Pomeron, and the leptons. 
The $Q^2$ evolution of most of these sets is based
on interpolation of a grid.
A larger selection of PDFs can be obtained via 
the interfaces to the \textsc{LHAPDF} libraries, one to the older 
Fortran-based \textsc{LHAPDF5} \cite{Whalley:2005nh} and one to the 
newer C++-based \textsc{LHAPDF6} \cite{Buckley:2014xyz}. 

Given that the 
\textsc{Pythia} machinery basically is a leading-order (LO) one, 
preference has been given to implementing LO sets internally.
In a LO framework, the PDFs have a clear physical interpretation 
as the number density of
partons, and can be related directly to measurable quantities.
In the modeling of minimum bias (MB) and underlying event (UE) phenomena, 
very small $x$ scales are probed, down to around $10^{-8}$, for $Q$ 
scales that may go below 1 GeV. 
Measurements of $F_2$ imply a 
small-$x$ behaviour for gluon and sea quark PDFs 
where $xf_i(x,Q^2)$ is constant or even slowly rising for $x \to 0$  
at a fixed $Q^2$ around $1-4~\mathrm{GeV}^2$.
This behaviour is evident in LO PDF fits.
Next-to-leading-order (NLO) PDFs, on the other hand,
no longer have a probabilistic interpretation, and their behaviour is
less directly related to physical quantities.
They
have small-$x$ corrections proportional to $\ln(1/x)$, that may drive
PDFs negative at small $x$ and $Q$.
This makes them unsuitable for describing showers or MPIs. 

Contrary to this argument, event tunes have been produced with NLO PDFs that
give a reasonable description of available collider data.
However, this is likely related to the resiliency of the 
MPI and string fragmentation frameworks, which allow 
a rather significant change of PDF shape to be
compensated by a retuning of relevant parameters. 
What is notable is that these NLO tunes require a 
significantly smaller $\pTo$ scale,
where $\pTo$ is used to tame the $1/\pT^4$ divergence of the 
QCD cross sections to $1/(\pT^2 + \pTo^2)^2$.
This reduced $\pTo$ 
compensates for the low amount of small-$x$ gluons in NLO PDFs. 
Since the integrated QCD cross
sections depend on the number density $f_i(x,Q^2)$, 
the small-$x$ partons play an important role in determining the
number and kinematics of the MPIs.
In the NLO tunes, the MPI collisions would tend to be symmetric, 
i.e.\ $x_1 \sim x_2$, and both not too small. 
Asymmetric collisions, where one $x$ is small, would be suppressed
by
the respective NLO PDFs vanishing or at least being tiny 
there (a negative PDF is reset to 0 in \textsc{Pythia}). 
With further scrutiny, one expects to find differences in the 
rapidity spectrum of minijets from MPIs.  Irrespective of that, 
there is no reason
to use NLO PDFs in regions where they are known not to be trustworthy.

If one is not satisfied to use a LO PDF set throughout, 
\textsc{Pythia} offers the possibility
to use two separate PDF sets; one
for the hard interaction and one
for the subsequent showers and MPI.  The former 
could well be chosen to be NLO and the latter LO. 
Recall, also, that ISR generated with the
standard backwards evolution scheme is based on ratios of PDFs. 
Therefore many of the differences between PDF sets divide out, 
notably away from the low-$x$ region.
An additional advantage of a two-PDF setup is 
that it becomes possible to explore a
range of PDFs for the hard process without any necessity 
to redo the UE/MB tune.

Some PDF studies in 
the \textsc{Pythia} context are found in \cite{Kasemets:2010sg}.   

\subsection{Parton showers}

The ISR and FSR algorithms are based on the dipole-style $\pT$-ordered
evolution first introduced in \textsc{Pythia}~6.3
\cite{Sjostrand:2004ef}. New features in \textsc{Pythia}~8 include 
$\gamma\to\qrm\qbar$ and $\gamma\to\ell^+\ell^-$ branchings as
part of the FSR 
machinery, options for emission of weak gauge  
bosons, $\Zrm^0$ and $\Wrm^{\pm}$, as part of both ISR and FSR
\cite{Christiansen:2014kba}, extensions of the shower
framework to handle bremsstrahlung in Hidden Valley
models~\cite{Carloni:2010tw,Carloni:2011kk}, and flexible colour
strengths for FSR when the emission rate is to be shared between
several recoilers \cite{Desai:2011su}, used e.g.\ to describe
radiation in R-parity violating SUSY decays. Options also
exist for alternative FSR recoil schemes and for 
gluon emissions off colour-octet onium states.
  
The entire perturbative evolution (ISR, FSR, and MPI) is interleaved into
a single common sequence of decreasing $\pT$ \cite{Corke:2010yf}.
The full interleaving performed in \textsc{Pythia}~8, which can be
switched on/off for cross checks, has
the beneficial consequence that the phase space available to hard FSR
emissions cannot end up depending on dipoles created by soft ISR ones,
hence we regard the new algorithm as more theoretically
consistent. 
Instead, some FSR is associated with colour dipoles 
stretched between a final-state parton and the beam-remnant "hole" left 
by an initial-state one, which therefore now can take a recoil. In the
ISR algorithm, recoils are always taken by the hard-scattering 
subsystem as a whole, regardless of whether the colour partner is in
the initial or final state. 

The shower evolution is based on the standard (LO) DGLAP splitting
kernels, $P(z)$~\cite{Gribov:1972ri,Altarelli:1977zs,Dokshitzer:1977sg}:
\begin{eqnarray}
P_{\qrm \to \qrm \grm}(z) & = & C_F\, \frac{1+ z^2}{1-z}~,\\
P_{\grm \to \grm \grm}(z) & = & C_A\, \frac{(1-z(1-z))^2}{z(1-z)}~,\\
P_{\grm\to \qrm \bar{\qrm}}(z) & = & T_R\, (z^2 + (1-z)^2)~, \label{eq:Pg2qq}
\end{eqnarray}
with $C_F=\frac43$, $C_A=N_C=3$, and $T_R=\frac12$, multiplied by $N_\frm$ 
if summing over all contributing quark flavours, for QCD, and
\begin{eqnarray}
P_{\frm \to \frm \gamma}(z) & = & e_{\frm}^2 \, \frac{1+ z^2}{1-z}~,\\
P_{\gamma\to \frm \bar{\frm}}(z) & = & e_{\frm}^2 \, N_C \, (z^2 + (1-z)^2)~,
\label{eq:Pgam2ff}
\end{eqnarray}
for QED, with $N_C = 1$ for charged leptons, and with $z$ the 
energy-sharing fraction between the daughter
partons. In addition, the current default is that gluon-polarisation
effects are taken approximately into account via a non-isotropic selection of 
the azimuthal angle of the branchings, $\varphi$. Corrections for
parton masses are generally also included, for both
FSR~\cite{Norrbin:2000uu} and ISR~\cite{Sjostrand:2004ef}. Additional
options for mass corrections for $\grm/\gamma \to \frm \bar{\frm}$
branchings are discussed below.

The ISR and FSR algorithms are both based on the above splitting
kernels, and are cast as differential equations expressing 
the probability of emitting radiation as one moves from 
high to low values of the shower evolution variable, which plays the
role of factorisation scale in parton-shower contexts. 
For FSR, this corresponds to an evolution forwards in physical time,
with a single mother parton replaced by two daughter
partons at each branching. For ISR, however, the progress from high
to low factorisation scales corresponds to a backwards evolution
in physical time~\cite{Sjostrand:1985xi}, with the evolving 
parton becoming unresolved into a new initial-state mother parton
and an accompanying final-state sister one at each
branching. Moreover, the fact that the
boundary condition represented by the non-perturbative structure of
the initial beam particle sits at the low-$Q$ end of the evolution
chain implies that a ratio of PDFs accompanies each branching, with the
purpose, roughly, of translating
from the PDF of the "old" mother parton to that of the
"new" one. 

Integrated over the kinematically allowed range of $z$ and 
expressed as a differential branching probability per unit evolution
time, the FSR and ISR kernels used to drive the shower evolution in
\textsc{Pythia} are:
\begin{eqnarray}
\frac{\drm {\cal P}_{\mathrm{FSR}}}{\drm  p_{\perp}^2} & = & \frac{1}{p_{\perp}^2}
\, \int \!\drm z\,\frac{\alphas}{2\pi}\, P(z)~, \nonumber\\[2mm]
\frac{\drm {\cal P}_\mathrm{ISR}}{\drm
  p_{\perp}^2} & = &
\frac{1}{p_{\perp}^2}\,
\int \!\drm z\, \frac{\alphas}{2\pi}\, P(z)\,
\frac{f'(x/z,p_{\perp}^2)}{zf(x,p_{\perp}^2)}
~,\label{eq:ISRFSRevol}
\end{eqnarray}
with $z=x/x'$ for ISR defined so $x<x'$, and 
the \textsc{Pythia} transverse-momentum evolution variable defined by
\begin{equation}
 p_\perp^2 ~=~p_{\perp\mathrm{evol}}^2~=~\left\{ 
 \begin{array}{lcl}
   (1-z)Q^2 & : & \mathrm{ISR}\\[2mm] 
   z(1-z)Q^2 & : & \mathrm{FSR}\\[2mm] 
 \end{array}
\right.~,\label{eq:pTevol}
\end{equation}
with $Q^2$ the offshellness of the branching parton: 
$Q_{\mathrm{FSR}}^2 = (p^2 - m_0^2)$ and $Q^2_\mathrm{ISR} = (-p^2 +
m_0^2)$, determined for each branching by solving the above equation
for $Q^2$. Note that, since FSR branchings involve timelike
virtualities ($p^2>0$) while ISR ones involve spacelike virtualities
($p^2<0$), both $Q^2$ definitions correspond to positive-definite
quantities.   

The overall
strength of radiation is set by the effective value of
$\alphas(M_\Zrm)$, which can be specified separately for ISR and 
FSR. Although nature
contains only a single strong coupling, the 
structure of higher-order splitting kernels differs between ISR
and FSR, and a further subtlety is that ISR involves an interplay with
PDFs, while FSR does not. Hence we believe there is ample
justification for maintaining two distinct effective $\alphas(M_\Zrm)$
values for bremsstrahlung, in addition to the ones which
govern hard processes and MPI.  

The default renormalisation scale used to
evaluate $\alphas$ for each shower branching is the shower evolution scale,
$p_{\perp\mathrm{evol}}$. 
For gluon-emission processes, this is the canonical choice of renormalisation
scale~\cite{Amati:1980ch},  
and it yields the correct ${\cal
  O}(\alpha^2_s\ln^3)$ behaviour at the integrated level. 
Optionally, a multiplicative prefactor can be applied,
$\mu_R^2 \,= \,k_{\mu_R} 
\,p_{\perp\mathrm{evol}}^2$, with default value $k_{\mu_R} = 1$. This can
be varied, e.g., to perform uncertainty estimates. 

We emphasise that the $\alphas(M_\Zrm)$ values used for ISR
and FSR in \textsc{Pythia} are not directly comparable to the
$\MSbar$ $\alphas(M_\Zrm)=0.1185(6)$ given by the
PDG~\cite{Beringer:1900zz}, for two reasons. 
Firstly, 
in the limit of soft-gluon emission ($z\to 1$), it can be
shown that the dominant ${\mathcal O}(\alphas^2)$ splitting-function term,
which generates contributions starting from ${\cal
  O}(\alphas^2\ln^2)$ at the integrated level, 
can be absorbed into the LO splitting functions by translating to the 
so-called CMW (a.k.a.~MC) scheme~\cite{Catani:1990rr},  
\begin{equation}
\alphas^{\mathrm{MC}} = \alphas^{\MSbar}\left(1 +
K\frac{\alphas}{2\pi}\right)~~~\longleftrightarrow~~~
\alphas^{\MSbar} = \alphas^{\mathrm{MC}}\left(1 - 
K\frac{\alphas}{2\pi}\right)~,
\end{equation}
where the scheme choice for the $\alphas/2\pi$ correction terms
amounts to an ${\cal   O}(\alphas^3)$ effect, and 
\begin{equation}
K \ = \ C_A\left(\frac{67}{18}-\frac16\pi^2\right)-\frac59
N_\frm \ = \
\left\{
\begin{array}{lcl}
4.565 & : & N_\frm = 3 \\
4.010 & : & N_\frm = 4 \\
3.454 & : & N_\frm = 5 \\
2.899 & : & N_\frm = 6
\end{array}
\right.~,\label{eq:KCMW}
\end{equation}
with $C_A=3$ and $N_\frm$ the number of contributing quark flavours.
The baseline value to use for shower $\alphas(M_\Zrm)$ values should
therefore be around $\alphas(M_\Zrm)^{\mathrm{MC}} = 0.126$. 
Secondly, even this larger value of $\alphas$ only takes into account
the NLO correction to the splitting kernel in the infinitely soft
limit. In the rest of phase space, the remaining NLO corrections still
tend to be positive, see e.g.~\cite{Hartgring:2013jma}, and hence the
effective value of $\alphas(M_\Zrm)$, when tuned directly to data, tends
to be a further 10\% larger, at $\alphas(M_\Zrm)^\mathrm{PYTHIA}\sim 0.139$.

In \textsc{Pythia}, one has the option of letting the translation
between the $\MSbar$ and MC schemes 
be done automatically, though the default is just to provide an
effective $\alphas(M_\Zrm)$ value directly in the \textsc{Pythia}
scheme. We also note
that the arguments in \cite{Catani:1990rr} were based on 2-loop
running, while the default in \textsc{Pythia} is to use 1-loop
running, which gives lower $\Lambda_\mathrm{QCD}$ values, 
allowing lower shower cutoffs to be used. 

For $\grm\to \qrm\qbar$ and
$\gamma\to\frm\bar{\frm}$ splitting processes,
the alternative choice of using $\mu_R \propto
m_{\frm\bar{\frm}}$ has also recently been implemented,
along with several options for the handling of mass 
effects, notably for charm and bottom quarks. The default now is
to start out from a splitting kernel,
\begin{equation}
P(z) = z^2 + (1-z)^2 + 8r_\frm z(1-z)~,
\end{equation}
normalised so that the z-integrated rate is $(\beta_\frm/3) (1 + r_\frm/2)$,
with $r_\frm = m_\frm^2/m_{\frm\bar{\frm}}^2$ and 
$\beta_\frm = \sqrt{1 - 4r_\frm}$, which should be the correct 
infinite-energy expression. This is then modified for finite dipole 
masses by an $(1 - m_{\frm\bar{\frm}}^2/m_{\rm dipole}^2)^3$ suppression 
factor, a factor which is derived from the 
$\Hrm^0 \to \grm \grm \to \qrm \qbar \grm$ matrix element,
but should be a reasonable estimate also for other processes.  

Concerning the emission of hard extra jets, one should be aware that 
the parton-shower machinery is primarily intended to describe
physics near the collinear and/or soft limits, in which successive
radiation $p_\perp$ scales are strongly ordered. 
Nevertheless, an extensive set of automated 
matrix-element corrections have been implemented, which correct the
first jet emission to the full LO matrix-element expression for a wide 
range of production and decay processes, see \cite{Miu:1998ju,Norrbin:2000uu}. 
For these processes, \textsc{Pythia} is therefore expected to achieve LO 
accuracy out of the box also for hard radiation.
For the FSR algorithm
such corrections are applied by default to all $1\to 2$ decay
processes in the SM and many BSM ones~\cite{Norrbin:2000uu}. For the
ISR shower, internal ME corrections have so far only been implemented
for radiation in a few hard $2\to 1$ processes, specifically 
$(\Zrm/\Wrm/\Hrm)+\mathrm{jet}$~\cite{Miu:1998ju}. For all other processes, 
an approximate improved-shower description is used to ensure a reasonable
behaviour up to the kinematic limit at high $\pT$ scales
\cite{Corke:2010zj}. Alternatively, see subsection
\ref{subsec:matchmerge} below for information on matching and merging
using external ME generators.

Finally, in the context of uncertainty estimates, it is worth noting
that the $\MSbar\to \mathrm{MC}$ scheme translation
is equivalent to making a specific shift of renormalisation scale, 
$\mu_R \to \mu_R \exp(-K/4\pi\beta_0) \sim \mu_R / 1.6$ (for $N_\frm=5$), 
with $\beta_0 = (11C_A - 2 N_\frm)/(12\pi)$ the 1-loop beta function in
QCD and $K$ defined by eq.~(\ref{eq:KCMW}). Therefore, making
arbitrary variations of $\mu_R$ around this 
scale will actually spoil the NLL precision of the shower,
at least in the infinitely soft limit in which the translation is
derived. So far, \textsc{Pythia} does not automatically attempt to
compensate for this, leaving it up to the user to judge which
variations to consider reasonable. 

\subsection{Multiparton interactions \label{sec:mpi}}

In hadron--hadron collisions, MPI are a
natural consequence of the composite structure of the colliding beam
particles. Although MPI are especially relevant to describe the
ubiquitous soft underlying event, the possibility of having 
several hard scattering processes occurring in one and the same
hadron--hadron collision also exists, albeit at suppressed rates
relative to soft MPI.

The basic formalism underpinning the MPI modeling in \textsc{Pythia}
is described in \cite{Sjostrand:1987su} and spans both soft and hard
QCD MPI processes in a single unified framework. The current
implementation, summarised briefly below, 
further contains the additional refinements 
introduced since \textsc{Pythia}~6.3 \cite{Sjostrand:2004pf}, along
with a few new additions unique to \textsc{Pythia}~8. In particular,
the mix of MPI processes has been enlarged from covering only
partonic QCD $2\to 2$ scattering in \textsc{Pythia}~6 to also allowing
for multiple
$\gamma+\mathrm{jet}$ and $\gamma\gamma$ processes, 
colour-singlet and -octet charmonium and bottomonium production, 
 $s$-channel $\gamma$ exchange, and $t$-channel
$\gamma/\Zrm^0/\Wrm^\pm$ exchange. 
Note also that for dedicated studies of two low-rate processes in
coincidence, the user can now request two distinct hard interactions
in the same event, with further MPI occurring as usual. 
There are then no Sudakov factors included for these two
interactions, similarly to normal events with one hard interaction. 

The starting point for parton-based MPI models is the observation
that the $t$-channel propagators  and $\alphas$ factors appearing in
perturbative QCD $2\to2$ scattering diverge at low momentum transfers, 
\begin{equation}
\drm {\sigma}_{2\to 2} \propto \ 
\frac{g_s^4}{16\pi^2}\frac{\drm t }{ {t}^2} \ \sim \ 
 \alpha^2_s({p}_\perp^2)\frac{{\drm}{{p}_\perp^2}}{{p}_\perp^4}~, \label{eq:dpt4} 
\end{equation}
a behaviour further exacerbated by the abundance of 
low-$x$ partons that can be accessed at 
large hadronic $\sqrt{s}$.
At LHC energies, this parton--parton cross section, integrated 
from some fixed ${p}_{\perp\mathrm{min}}$ scale up to the kinematic maximum, 
becomes larger than the total hadron--hadron cross section for
${p}_{\perp\mathrm{min}}$ values of order $4-5~\mbox{GeV}$. 
In the context
of MPI models, this is interpreted straightforwardly 
to mean that \emph{each} hadron--hadron collision
contains \emph{several} parton--parton collisions, with typical 
momentum transfers of the latter of order $p_{\perp\mathrm{min}}$. 

This simple reinterpretation in fact expresses unitarity;
instead of the total interaction cross
section diverging as $p_{\perp\mathrm{min}} \to 0$, which would violate
unitarity, we have restated the problem so
that it is now the \emph{number of MPI per collision} that
diverges, while the total cross section remains finite. 

Taking effects beyond (unitarised) $2\to 2$ perturbation theory into account, 
the rise of the parton--parton cross section for $p_\perp\to 0$ must
ultimately be tamed by colour-screening effects; the individual
coloured constituents of hadrons cannot be resolved by infinitely long
(transverse) wavelengths, analogously to how hadronisation provides a
natural lower cutoff for the perturbative parton-shower evolution.
In \textsc{Pythia}, rather 
than attempting an explicit dynamical modeling of screening and/or 
saturation effects, this aspect is implemented via the effective replacement, 
\begin{equation}
\frac{\drm {\sigma}_{2\to 2}}{\drm {p}_{\perp}^2} \propto
\frac{\alphas^2({p}_\perp^2)}{{p}_\perp^4} \rightarrow
\frac{\alphas^2({p}_\perp^2 + p_{\perp 0}^2)}{({p}_\perp^2 +
  p_{\perp 0}^2)^2}~,\label{eq:sigmaPT0}
\end{equation}
which smoothly regulates the divergence. The MPI cross section in
the $p_\perp \to 0$ limit thus tends to a constant, the size of which
is controlled directly by: 
\begin{enumerate}
\item the effective $p_{\perp 0}$ parameter, 
\item the value of $\alphas(M_\Zrm)$ used for MPI and its
running order, and 
\item the PDF set used to provide the parton luminosities for MPI. 
\end{enumerate}
These are therefore the three main tunable
aspects of the model. Two further highly important ones are the
assumed shape of the hadron mass distribution in impact-parameter
space, and the strength and modeling of
colour-reconnection effects.

To be more explicit, 
the regulated parton--parton cross section,
eq.~(\ref{eq:sigmaPT0}), can be integrated to 
provide a first rough estimate of how many MPI, on average, occur in
each (average, inelastic non-diffractive) hadron--hadron collision, 
\begin{equation}
 \langle n_\mathrm{MPI} \rangle({p}_{\perp 0}) = \frac{\sigma_{2\to2}(p_{\perp 0})}
{\sigma_{\rm ND}} ~,\label{eq:nMPI}
\end{equation}
with $\sigma_\mathrm{ND}$ given by eq.~(\ref{eq:sigND}). 
This formula would only be strictly true if all the MPI
could be considered equivalent and independent, i.e.\ uncorrelated,  
in which case $\langle n_\mathrm{MPI} \rangle$ could be
interpreted as the mean of a Poisson distribution. In \textsc{Pythia},
this is not the case, since several correlation effects are taken into
account, some of which can be quite important, notably energy
conservation among the partons in each beam hadron.  This is achieved
by first reorganizing the MPI into  
an ordered sequence of falling $p_\perp$ values
\cite{Sjostrand:1987su}, similarly to what is done for perturbative 
bremsstrahlung emissions in the parton-shower formalism, 
so that the hardest MPI is generated first. 
The probability for an interaction, $i$, is then given by a
Sudakov-type expression, 
\begin{equation}
\frac{\drm \mathcal{P}_{\mathrm{MPI}}}{\drm {\pT}} =
\frac{1}{\sigma_{\mathrm{ND}}} \frac{\drm \sigma_{2\to 2}}{\drm \pT} \;
\exp \left( - \int_{\pT}^{p_{\perp i-1}}
\frac{1}{\sigma_{\mathrm{ND}}} \frac{\drm \sigma_{2\to 2}}{\drm \pT'} 
\drm \pT' \right),
\label{eq:MPIevol}
\end{equation}
where $\drm\sigma_{2\to 2}$ may now be modified to take correlations
with the $(i-1)$ preceding MPI into account. In particular, 
momentum conservation is achieved by ``squeezing'' the PDFs into the remaining
available $x$ range, while adjusting their normalisations to
respect number-counting sum rules~\cite{Sjostrand:1987su},
\begin{equation}
f_i(x) \to \frac{1}{X}f_0\left(\frac{x}{X}\right)~,
\end{equation}
with subscript $0$ referring to the original, one-parton inclusive
PDFs, and $X$ the momentum fraction remaining in the beam
remnant after the preceding $(i-1)$ interactions, including any subsequent
modifications to their $x$ fractions by ISR showering, 
\begin{equation}
X = 1-\sum_{m=1}^{i-1}x_m~.
\end{equation}
Flavour conservation is imposed by
accounting for how many of the preceding MPI involved valence and/or
sea quarks, so that the full forms of the PDFs used for the $i$'th
MPI are~\cite{Sjostrand:2004pf}: 
\begin{eqnarray}
f_{i}(x,Q^2) & = & \frac{N_{f\mathrm{v}}}{N_{f\mathrm{v} 0}} \frac{1}{X}
f_{\mathrm{v}0}\left(\frac{x}{X},Q^2\right)
 + \frac{a}{X} f_{\mathrm{s} 0}\left(\frac{x}{X},Q^2\right)
 + \sum_j \frac{1}{X} f_{\mathrm{c}_j 0}\left(\frac{x}{X};x_j\right)\\[2mm]
g_i(x,Q^2) & = & \frac{a}{X}g_0(x,Q^2) ~,
\end{eqnarray}
with $f_i(x, Q^2)$ ($g_i(x, Q^2)$) being the squeezed PDFs for quarks (gluons), 
$N_{f\mathrm{v}}$ ($N_{f\mathrm{v} 0}$) the number of remaining (original) 
valence quarks of the given flavour in the beam remnant, $f_{\mathrm{s}}$ the
sea-quark PDF, $f_{\mathrm{c}_j}$ a so-called \emph{companion
PDF} derived from $\grm\to \qrm\qbar$ splitting whenever a sea quark
($j$) is kicked out, 
and the common normalisation factor for the gluons + sea-quarks, $a$,
defined to satisfy the total momentum sum
rule~\cite{Sjostrand:2004pf}. While this is still less than a    
full multi-parton QCD evolution, it has the advantages of remaining 
straightforward to work with for arbitrarily many MPI initiators,
preserving the endpoint behaviours for $x\to X$, and, 
at the very least, it obeys the momentum and flavour sum rules
explicitly, hence we expect the dominant such correlations to be
included in the formalism. 

A further aspect of the MPI picture is the impact-parameter dependence. 
Protons are extended objects, and thus collisions may vary from central 
to peripheral. The more central, the bigger the overlap between the
colliding cloud of partons, and the larger the average number of MPIs
per collision. The shape of the proton thus makes a difference.
The more uneven this distribution, i.e.\ the more sharply peaked it is 
in the middle, the easier it is for a central collision to yield a large
number of MPIs and thereby a large charged multiplicity. With the RMS
spread (approximately) given by the measured proton radius, a more
sharply peaked distribution also has longer low-level tails, giving
more low-multiplicity events. The width of the multiplicity distribution
therefore is a good indicator of the partonic distribution inside the
proton, even if it is influenced by other
contributing factors. \textsc{Pythia} implements several alternative
shapes that can be compared. The simplest is a Gaussian profile
--- very convenient for convolutions of the two incoming hadrons ---
that does not appear to be too far off from what is needed to describe 
data, even if best tunes typically are obtained with distributions
somewhat more uneven than this. 

As already mentioned, MPI is now combined with ISR and FSR to provide one
common sequence of interleaved $\pT$-ordered interactions or branchings
\cite{Corke:2010yf}, defined by 
\begin{eqnarray}
\frac{\drm \mathcal{P}}{\drm \pT} & = & 
\left( \frac{\drm\mathcal{P}_{\mathrm{MPI}}}{\drm \pT}  + 
\sum   \frac{\drm\mathcal{P}_{\mathrm{ISR}}}{\drm \pT}  +
\sum   \frac{\drm\mathcal{P}_{\mathrm{FSR}}}{\drm \pT} \right)
\nonumber \\ 
 & \times & \exp \left( - \int_{\pT}^{p_{\perp i-1}} 
\left( \frac{\drm\mathcal{P}_{\mathrm{MPI}}}{\drm \pT'}  + 
\sum   \frac{\drm\mathcal{P}_{\mathrm{ISR}}}{\drm \pT'}  +
\sum   \frac{\drm\mathcal{P}_{\mathrm{FSR}}}{\drm \pT'} \right) 
\drm \pT' \right) ~,
\label{eq:combinedevol}
\end{eqnarray}
where $p_{\perp i-1}$ is the $\pT$ scale of the previous
step, the FSR and ISR evolution kernels given by
eq.~(\ref{eq:ISRFSRevol}), and the MPI one by eq.~(\ref{eq:MPIevol}). 
This thus constitutes the \emph{master evolution
equation} of \textsc{Pythia}~8.  

Finally, two additional and optional new components (off by default) 
are also available in \textsc{Pythia}~8: 
\begin{enumerate}
\item a model for partonic rescattering, i.e.\ that an outgoing parton 
from one interaction can be incoming to another~\cite{Corke:2009tk}, and 
\item an option for an $x$-dependent impact-parameter shape, 
where high-momentum partons are located closer to the center of the
hadron than low-momentum ones~\cite{Corke:2011yy}.
\end{enumerate}

\subsection{Beam remnants and colour reconnection}

The extraction of several MPI initiators from the incoming hadrons can 
leave behind quite complicated beam remnants, potentially in high
colour representations. In the default beam-remnant model, gluon 
initiators are attached to other colour lines so as to reduce the 
total colour charge associated with the remnant, short of making it a
singlet \cite{Sjostrand:2004pf}. A new option allows for arbitrary
colour representations, with a flexible suppression of higher-charged   
states \cite{Christiansen:2014xyz}. 

Each initiator parton should have a certain Fermi motion inside the 
hadron, \emph{primordial $k_{\perp}$}, expected to be a few hundred MeV. 
In the study of observables such as the $\pT$ spectrum of the $\Zrm$ 
gauge boson in the low end of the distribution, average values around 
2~GeV instead are preferred. Likely such a high value reflects low-$\pT$ 
ISR branchings that are not fully simulated; recall that a low-$\pT$ 
cut-off on branchings is imposed because the emission rate diverges 
and therefore becomes unmanageable. In the code, a reasonably flexible 
ansatz is used wherein the width of the primordial $k_{\perp}$ distribution can 
depend on the scale of the hard process itself, so that low-$\pT$ MPI 
systems do not have as large a primordial $k_{\perp}$ as high-$\pT$ ones.   

Data suggest the existence of colour reconnection \cite{Sjostrand:1987su}, 
whereby the colour flow of the different MPIs get mixed up, over and above 
what is already implied by the beam-remnant model. Currently three models 
are implemented in the \textsc{Pythia} core library.
\begin{itemize}
\item The \emph{MPI-based model}, which is the original and default option,
wherein all the gluons of a lower-$\pT$ interactions can be inserted
onto the colour-flow dipoles of a higher-$\pT$ one, in such a way
as to minimise the total string length \cite{Argyropoulos:2014zoa}.
\item The \emph{QCD-based model}, wherein alternative coherent
  parton--parton states beyond leading colour are identified based on
  the multiplet structure of $SU(3)_C$, and reconnections are
  allowed to occur when the total string length can be reduced
  \cite{Christiansen:2014xyz}. Particular attention is  
given to the formation of \emph{junctions}, i.e.\ where three string pieces 
form a Y-shaped topology, which provides an additional source of
baryon formation in this model.
\item The \emph{gluon-move model}, wherein individual gluons are moved from
their current location, on the colour line in between two partons,
to another such location if that results in a reduction of the total 
string length \cite{Argyropoulos:2014zoa}. An optional ``flip'' step
can reconnect two different string systems, such that a quark end 
becomes connected with a different antiquark one.
\end{itemize} 
A further selection of models \cite{Argyropoulos:2014zoa} is available,
but only as less well supported plugins. The hadron-collider models from 
\textsc{Pythia}~6 \cite{Skands:2007zg} have not been implemented, 
and neither (yet) the $\Wrm^+ \Wrm^-$ machinery studied at
LEP \cite{Sjostrand:1993hi}.

\subsection{Hadronisation}

Hadronisation ---  the mechanism for transforming the final outgoing 
coloured partons into colourless particles --- is based solely on the 
Lund string fragmentation framework \cite{Andersson:1983ia,Sjostrand:1984ic}; 
older alternative descriptions have been left out. The handling of 
junction topologies has been improved, allowing more complicated 
multijunction string configurations \cite{Christiansen:2014xyz}, but the 
core string fragmentation machinery remains the same since many years. 
Historically it is at the origin of the \textsc{Jetset/Pythia} programs.

While non-perturbative QCD is not solved, hadron spectroscopy and
lattice QCD studies lend support to a linear confinement picture in the 
absence of dynamical quarks, i.e.\ the energy stored in the colour
dipole field between a charge and an anticharge increases linearly
with the separation between the charges, if the short-distance
Coulomb term is neglected. The assumption of linear confinement
provides the starting point for the string model, most easily
illustrated for the production of a back-to-back $\qrm\qbar$ jet pair.
As the partons move apart, the physical picture is that of a colour
flux tube, or maybe colour vortex line, being stretched between 
the $\qrm$ and the $\qbar$. The transverse dimensions of the tube are 
of typical hadronic sizes, roughly 1~fm, and the string tension, i.e.\ 
the amount of energy per unit length, is $\kappa \approx 1$~GeV/fm. 
In order to obtain a Lorentz covariant and causal description of the 
energy flow due to this linear confinement, the most straightforward way 
is to use the dynamics of the massless relativistic string with no 
transverse degrees of freedom. The mathematical, one-dimensional string 
can be thought of as parametrizing the position of the axis of a 
cylindrically symmetric flux tube. 

As the $\qrm$ and $\qbar$ move apart, the potential energy stored 
in the string increases, and the string may break by the production of 
a new $\qrm' \qbar'$ pair, so that the system splits into two colour 
singlet systems $\qrm \qbar'$ and $\qrm' \qbar$. If the invariant mass 
of either of these string pieces is large enough, further breaks occur 
until only on-shell hadrons remain, each hadron corresponding to 
a small piece of string.
 
In general, the different string breaks are causally disconnected.
This means that it is possible to describe the breaks in any convenient
order, e.g.\ from the quark end inwards, and also include as constraint 
that the hadrons produced must have their physical masses. Results, 
at least not too close to the string endpoints, should be the same if 
the process is described from the $\qrm$ end or from the $\qbar$ one. 
This left--right symmetry constrains the allowed shape of fragmentation 
functions, $f(z)$, which describe how energy is shared between the hadrons.
The shape contains some free parameters, however, which have to be 
determined from data.

The flavour composition of the new quark--antiquark pairs $\qrm' \qbar'$ 
is assumed to derive from a quantum mechanical tunneling process.
This implies a suppression of heavy quark production,
$\mathrm{u} : \mathrm{d} : \mathrm{s} : \mathrm{c} \approx 
1 : 1 : 0.3 : 10^{-11}$, such that charm and bottom production can be 
neglected in the hadronisation step. Tunneling also leads to a 
flavour-independent Gaussian spectrum for the transverse momentum of 
$\qrm' \qbar'$ pairs. A tunneling mechanism can also be used to explain 
the production of baryons, but this is still a poorly understood area. 
In the simplest possible approach, a diquark in a colour antitriplet 
state is just treated like an ordinary antiquark, but it is also 
possible to imagine sequential production of several $\qrm \qbar$ 
pairs that subsequently combine into hadrons, the so-called 
\emph{popcorn} model \cite{Andersson:1984af}.
 
If several partons are moving apart from a common origin, the details 
of the string drawing become more complicated. For a $\qrm \qbar \grm$ 
event, a string is stretched from the $\qrm$ end via the $\grm$ to the 
$\qbar$ end, i.e.\ the gluon is a kink on the string, carrying energy and 
momentum. As a consequence, the gluon has two string pieces attached, 
and the ratio of gluon/quark string forces is 2, a number that
can be compared with the ratio of colour charge Casimir operators,
$N_C/C_F = 2/(1-1/N_C^2) = 9/4$. In this, as in other
respects, the string model can be viewed as a variant of QCD,
where the number of colours $N_C$ is not $3$ but infinite.
Fragmentation along this kinked string proceeds along the same lines
as sketched for a single straight string piece. Therefore no new 
fragmentation parameters have to be introduced, a most economical aspect
of the model.

In a high-energy event hundreds of partons may be produced, so the 
string topology becomes quite complicated. Colours are book-kept, 
in the $N_C \to \infty$ limit, for the selection of hard processes 
and in shower branchings. The colour flows in separate MPIs become 
correlated via the beam remnants, and colour reconnection can move 
colours around. At the end of the day the colours can be traced,
and the event may subdivide into a set of separate colour singlets,
as follows. Each open string has a colour triplet (a quark or an 
antidiquark) at one end, an antitriplet at the other, and a number of 
gluons in between. A closed string corresponds to a ring of connected 
gluons. 

A further component is the junction~\cite{Sjostrand:2002ip} of three 
string pieces in a Y-shaped topology. With each piece ending at a quark, 
the junction comes to be associated with the net baryon number of the 
system. An antijunction similarly is associated with a antibaryon number. 
In general, a system can contain several junctions and antijunctions,
and then the description can become quite unwieldy. Typically 
simplifications are attempted, wherein the big system is split into 
smaller ones, each containing one (anti)junction. 

\subsection{Resonance and particle decays}

In \textsc{Pythia} a technical distinction is made between the following 
terms:
\begin{itemize}
\item resonances: states with a typical lifetime shorter 
than the hadronisation scale,
\item particles: states with a lifetime comparable to 
or longer than the hadronisation scale, and
\item partons: states with colour which must be
hadronised. 
\end{itemize}
In practical terms, any state with an on-shell mass above
$20$~GeV in \textsc{Pythia} is by default treated as a resonance,
e.g.\ $\gamma^*/\Zrm^0$, $\Wrm^\pm$, top, Higgs bosons, and most
BSM states such as 
sfermions and gauginos. However, some light hypothetical weakly
interacting or stable states such as the gravitino are also considered
as resonances. All remaining colourless states, primarily leptons and
hadrons, are treated as particles, while quarks and gluons are
partons.

All resonances are decayed sequentially as part of the hard process,
and so the total cross-section as calculated by \textsc{Pythia} is
dependent upon the available decay channels of the resonance. Closing
a channel will decrease the cross-section accordingly. Conversely,
particle decays are performed after hadronisation, and changing the
decay channels of a particle will not affect the total
cross-section. It is important to note that in this scheme states such
as the $\rho$, $\Jrm/\psi$, and $\Upsilon$ are considered particles and
not resonances. Consequently, allowing only the decay $\Jrm/\psi \to
\mu^+\mu^-$ does not change the cross-section for the hard process
$\grm \grm \to \Jrm/\psi \grm$. The rationale here is that particles, such as
the $\Jrm/\psi$, can also be produced by parton showers, string fragmentation 
and particle decays, e.g.\ $\grm \to \brm\bbar$, $\bbar \to \Brm$, 
$\Brm \to \Jrm/\psi$. Any bias at the hard-process level would not affect
these other production mechanisms and thus be misleading rather than helpful.
Instead the user must consider all
relevant production sources and perform careful bookkeeping. 

By default, particles are decayed isotropically. However, many particle
decays are then weighted using generic matrix elements, such as for a
Dalitz decay or a weak decay. Additionally, the handling
of tau decays has been significantly improved
\cite{Ilten:2012zb,Ilten:2014jva}. Notably, full spin correlations are
calculated for tau decays produced from most standard mechanisms
including all EW production, Higgs production, and production
from $\Brm$- and $\Drm$-hadrons. Taus can also be decayed using
polarisation information passed from LHE files.

Tau spin correlations are calculated using the helicity density formalism,
where the weight for an $n$-body tau decay is given by,
\begin{equation}
  \mathcal{W} = 
  \rho_{\lambda_0\lambda_0'}
  \mathcal{M}_{\lambda_0;\lambda_1 \ldots \lambda_n}
  \mathcal{M}_{\lambda_0';\lambda_1' \ldots \lambda_n'}^*
  \prod_{i = 1,n} \mathcal{D}_{\lambda_i\lambda_i'}^{(i)}~.
  \label{eq:decayW}
\end{equation}
Here, the tau is indexed by $0$ and its decay products with $1$
through $n$, where the helicity for the $i^\mathrm{th}$ particle is
given by $\lambda_i$; repeated helicity indices are summed over. The
full helicity density matrix for the tau is $\rho$, while the helicity
dependent matrix element for the decay is $\mathcal{M}$, and the decay
matrix for each outgoing particle is $\mathcal{D}$.

The helicity density matrix for the $i^\mathrm{th}$ outgoing particle
in a $2 \to n$ process is calculated by
\begin{equation}
  \rho_{\lambda_i \lambda_i'}^{(i)} = 
  \rho_{\kappa_1\kappa_1'}^{(1)}\rho_{\kappa_2\kappa_2'}^{(2)}
  \mathcal{M}_{\kappa_1\kappa_2; \lambda_1 \ldots \lambda_n}
  \mathcal{M}_{\kappa_1'\kappa_2'; \lambda_1' \ldots \lambda_n'}^*
  \prod_{j \neq i} \mathcal{D}_{\lambda_j\lambda_j'}^{(j)}~,
  \label{eq:decayRHO}
\end{equation}
where $\rho^{(1,2)}$ are the helicity density matrices for the
incoming particles and $\mathcal{M}$ is the helicity matrix element
for the process. For incoming two-helicity-state particles with a
known longitudinal polarisation $\mathcal{P}_z$, e.g.\ beam particles,
the helicity density matrix is diagonal with elements
$(1\pm\mathcal{P}_z)/2$.

In eqs.~(\ref{eq:decayW}) and~(\ref{eq:decayRHO}), if no particles in
the chain have been decayed, all decay matrices $\mathcal{D}$ are
initially given by the identity matrix. However, if two taus are
produced from the same mechanism, then the decay matrix for the first
decayed tau is calculated with
\begin{equation}
  \mathcal{D}_{\lambda_0\lambda_0'} = 
  \mathcal{M}_{\lambda_0;\lambda_1 \ldots \lambda_n}
  \mathcal{M}_{\lambda_0';\lambda_1' \ldots \lambda_n'}^* \prod_{i
    = 1,n} \mathcal{D}_{\lambda_i\lambda_i'}^{(i)}~,
  \label{eq:decayD}
\end{equation}
and then used in eq.~(\ref{eq:decayRHO}) when calculating the helicity
density matrix for the second tau. In this way the decays of the two
taus are correlated.

Dedicated tau decay models, similar to those available in 
\textsc{Tauola}  \cite{Jadach:1993hs}, are
available for up to six-body tau decays and are provided for all decay
channels with branching fractions greater than $0.04\%$. The helicity
density matrix for each tau decay model, as used in eqs.~(\ref{eq:decayW})
and~(\ref{eq:decayD}) can be generalised as
\begin{equation}
  \mathcal{M} = \left(\frac{g_w^2}{8 m_W^2}\right) 
  \bar{u}_{\nu_\tau} \gamma_\mu (1 - \gamma^5) u_\tau J^\mu~.
\end{equation}
For the implementation of new tau decay models only the hadronic
current $J^\mu$ must be provided.

Most particle data (resonances, particles, and partons) have been
updated in agreement with the 2012 PDG tables
\cite{Beringer:1900zz}. This also includes a changed content of the
scalar meson multiplet. Some updated charm and bottom decay tables
have been obtained from the DELPHI and LHCb collaborations.

The BE$_{32}$ model for Bose--Einstein effects \cite{Lonnblad:1997kk} 
has been implemented, but is not operational by default.  

\subsection{Matching and merging \label{subsec:matchmerge}}

Matching and merging techniques attempt to provide a consistent
combination of a matrix-elements-based description at high 
momentum-transfer scales and a parton-shower-based one at low
scales \cite{Buckley:2011ms}. A wide variety of techniques have
been proposed over the years, and many of them are available in 
\textsc{Pythia}, usually requiring external ME 
events that then internally are accepted or rejected, and combined
with parton showers (PS).

Combinations of fixed order results with the parton shower 
resummation broadly fall into two categories.
Process-specific schemes provide improvements for specific physics processes.
Typically, this means a better description of the radiation pattern of the
first emission, often combined with an improved description of the inclusive
cross section. Improvements of this type in \textsc{Pythia} are first-order
ME corrections \cite{Norrbin:2000uu, Miu:1998ju}, POWHEG 
matching \cite{Nason:2004rx,Frixione:2007vw} and
MC@NLO matching \cite{Frixione:2002ik,Platzer:2011bc,
Hoeche:2011fd,Alwall:2014hca}.

Process-independent schemes attempt to supplement the parton shower with
multiple fixed-order calculations simultaneously, independent of the core
process $X$. This usually means that exclusive observables with $X+0,~X+1,\dots
X+N$ resolved partons are described with fixed-order accuracy. External inputs
are needed. These merging schemes typically separate
the phase space for emissions into hard and soft/collinear regions by means of
a jet criterion, and use the parton shower to fill soft/collinear emissions
while using the fixed-order calculation to provide hard emissions. Improvements
of this type in \textsc{Pythia} are CKKW--L multi-leg merging 
\cite{Lonnblad:2001iq, Lonnblad:2011xx}, 
MLM jet matching \cite{Mangano:2006rw},
unitarised ME+PS merging \cite{Lonnblad:2012ng}, NL$^3$ and unitarised
NLO+PS merging \cite{Lonnblad:2012ix}, and FxFx NLO jet matching
\cite{Frederix:2012ps}.

We will briefly introduce the improvements that are
available in \textsc{Pythia} below.

\subsubsection{Process-specific improvements}

The earliest process-specific improvements still used are first-order
ME corrections \cite{Bengtsson:1986hr}. Within these
schemes, the parton shower emission probability, which normally only
captures the singularities of real-emission corrections, 
is upgraded to reproduce the full $+1$-parton inclusive tree-level 
cross section. Such improvements exist for the resonance decays 
$\Zrm \to \qrm\qbar\grm$, $\Hrm \to \qrm\qbar\grm$, 
$\Wrm \to \qrm\qbar'\grm$, and $\trm \to \Wrm\brm\grm$ \cite{Norrbin:2000uu}, 
as well as the processes $\prm\prm \to\Zrm+$parton, 
$\prm\prm \to \Hrm+$parton and $\prm\prm \to\Wrm+$parton
\cite{Miu:1998ju}.

To see how this works, consider a lowest-order
$n$-body process. Define the phase space mapping $\drm\Phi_{\mathrm{rad}}$, 
whereby a shower radiation turns an $n$-body state into an $n+1$-body one,
i.e.\ $\drm\Phi_{n+1} = \drm\Phi_n \, \drm\Phi_{\mathrm{rad}}$. Let  
$B_n$ and $B_{n+1}$ denote the corresponding Born-level cross sections. 
Then an $n$-body event is first selected according to $B_n \drm\Phi_n$. 
Thereafter an emission is picked according to 
\begin{equation}
\frac{B_{n+1}}{B_n} \, \drm\Phi_{\mathrm{rad}} \, \, \Delta(Q_{\mathrm{max}}, Q)
~,~~~~\mathrm{where}~~~ \Delta(Q_{\mathrm{max}}, Q) =
\exp \left( - \int_Q^{Q_{\mathrm{max}}} 
\frac{B_{n+1}}{B_n} \, \drm\Phi_{\mathrm{rad}} \right) ~.
\label{eq:oldmatch}
\end{equation} 
Here $Q$ is the evolution variable of the parton shower, starting at a 
hard scale and evolving towards softer ones. Neglecting the $\Delta$ factor,
the original selection by $B_n$, multiplied by the $B_{n+1}/B_n$ emission 
rate, combines
to recover $n+1$-body production rate of $B_{n+1} \drm\Phi_{n+1}$. 
The exponential $\Delta(Q_{\mathrm{max}}, Q)$
introduces the standard Sudakov suppression, that a first
emission cannot occur at a lower scale if it already occured at a higher one.
Put another way, the $B_{n+1}/B_n$ ratio replaces the normal shower branching
probability, in the prefactor as well as in the Sudakov factor. Note 
that the $B_{n+1}/B_n$ ratio is divergent in the $Q \to 0$ soft/collinear 
limit. Thereby the first expression of eq.~(\ref{eq:oldmatch}) 
integrates to unity, i.e.\ the 
procedure formally always picks one emission. In practice a lower 
shower cutoff $Q_{\mathrm{min}}$ is introduced. Thereby the events that
reach this scale in the downwards evolution in $Q$ remain as $n$-body
events, while events with an emission above $Q_{\mathrm{min}}$ are 
$n+1$-body ones. Once a branching has been
chosen at a scale $Q$, an ordinary shower is allowed to start at this
scale and continue to evolve downwards. But the improved emission 
probabilities derived for these specific cases can also be used after 
the first emission, provided that the flavour structure does not change 
by a $\grm\to\qrm\qbar$ splitting, as the improved splitting kernels 
$B_{n+1}/B_n$ provide a better approximation of the QCD emission cross 
sections than the regular ones. 

NLO matching methods form a rather natural extension of ME  
corrections. The POWHEG method 
\cite{Nason:2004rx,Frixione:2007vw,Alioli:2008tz} 
in its original form follows the prescription
above, with one major and two minor differences. The major one is that
the original selection according to $B_n \drm\Phi_n$ is replaced by
$\overline{B}_n \drm\Phi_n$. Here $\overline{B}_n$ is the differential
NLO cross section, wherein the $B_{n+1} \drm\Phi_{n+1}$ rate is integrated 
over the $\drm\Phi_{\mathrm{rad}}$ emission phase space and combined with 
the virtual $n$-body corrections and $B_n$ itself. Starting from this,
eq.~(\ref{eq:oldmatch}) is used as before to provide exactly one hard
emission. The two minor differences are that the phase-space mapping
$\drm\Phi_{\mathrm{rad}}$ and the hardness ordering variable $Q$ are
likely to be different from those in standard showers. 

One consequence of the procedure is that, in the hard region where
the Sudakov is close to unity, the $B_{n+1}$ cross section is multiplied
by a \emph{$K$ factor} $\overline{B}_n/B_n$. It is possible to split 
$B_{n+1}$ into a soft and a hard piece, 
$B_{n+1} = B_{n+1}^{\mathrm{S}} + B_{n+1}^{\mathrm{H}}$, however, where only
the soft piece is rescaled by a $K$ factor. To this end a function 
$F(\Phi_{\mathrm{rad}})$ 
is introduced, with $B_{n+1}^{\mathrm{S}} = F(\Phi_{\mathrm{rad}}) B_{n+1}$ and
$B_{n+1}^{\mathrm{H}} = (1 - F(\Phi_{\mathrm{rad}})) B_{n+1}$. The only strict
requirement on the $F$ function is that it should approach unity
in the soft/collinear region, and sufficiently fast so that 
$\overline{B}_n^{\mathrm{S}} = \overline{B}_n 
- \int B_{n+1}^{\mathrm{H}} \drm\Phi_{\mathrm{rad}} > 0$. Events sampled
according to $\overline{B}_n \drm\Phi_n$ are classified as soft with a
probability $\overline{B}_n^{\mathrm{S}} / \overline{B}_n$, and else hard. 
For hard events an emission is 
picked by $B_{n+1}^{\mathrm{H}} \drm\Phi_{n+1}$, i.e.\ without any 
$K$ factor or Sudakov factor. For soft ones the formalism of 
eq.~(\ref{eq:oldmatch}) can be applied as before, only with 
$B_{n+1}^{\mathrm{S}}$ in place of  $B_{n+1}$.   

The \textsc{Powheg-Box} program \cite{Alioli:2010xd} calculates the NLO cross 
section and performs a first emission as described above, except for
the fraction of the event where the emission would be below the 
$Q_{\mathrm{min}}$ scale. Normally $Q_{\mathrm{min}}$ is chosen to be 
reasonably small, around 2~GeV, say, so that the no-emission fraction
of events is also small. These output events have to be interfaced to an
external event generator to achieve a complete NLO+PS event generation.
\textsc{Pythia} provides an interface to \textsc{Powheg-Box}-produced input 
files. The difficulty in interfacing LHEF inputs to the parton shower lies in 
avoiding overlaps: partonic states available through the POWHEG method 
should not be produced by subsequent parton showering. Had the shower
used exactly the same hardness criterion $Q$ as \textsc{Powheg-Box}, this would
have been trivial; continue the shower evolution downwards from this
$Q$ scale of the event or, if there was no emission, from the 
$Q_{\mathrm{min}}$ scale. The problem is that both \textsc{Pythia} and
\textsc{Powheg-Box} use transverse momentum, but somewhat differently defined.  
The solution is to use vetoed showers, where the shower is started off
from the maximum scale, but where all emissions that are above the  
\textsc{Powheg-Box} $Q$ scale are rejected. The LHEF~1.0 accord does transfer 
the $Q$ scale of the current event, but does not convey the functional
form of the phase-space boundary it implies. Therefore, \textsc{Pythia} 
supports various hardness definitions upon which to base vetoed showering.

The second major NLO matching method is the MC@NLO strategy
\cite{Frixione:2002ik}. In MC@NLO,  the NLO cross section
is again split into a soft $\overline{B}_n^{\mathrm{S}} $ and a hard part
$B_{n+1}^{\mathrm{H}}$. The difference is that here the soft part is
defined to
only include soft/collinear real emission contributions
$B_{n+1}^{\mathrm{S}}$,
as given by a specific parton shower, $B_{n+1}^{\mathrm{S}} = B_n
\otimes \mathcal{K}^{\mathrm{PS}}$. Here $\mathcal{K}^{\mathrm{PS}}$ 
is the shower splitting
kernel, i.e.\ omitting the Sudakov form factor. The soft part is bookkept
as an $n$-body configuration. The hard real radiation part excludes
soft contributions and is given by $B_{n+1}^{\mathrm{H}} = B_{n+1} -
B_{n+1}^{\mathrm{S}}$. Note that showers are expected to reproduce
the ME behaviour in the soft/collinear limit, such that both
$\overline{B}_n^{\mathrm{S}} $ and $B_{n+1}^{\mathrm{H}}$ are
separately finite. Away from this limit, it is an advantage if the shower
emission rate everywhere is below the ME one --- if not, negative-weight
events will have to be used.

The a\textsc{MC@NLO} program  \cite{Alwall:2014hca} automates the
generation of ME-level events for various common parton
showers, including \textsc{Pythia}. These are
stored in form of LHE files containing soft $n$-body events and hard
$n + 1$-body ones. The former are selected according to
$\overline{B}_n^{\mathrm{S}} = \overline{B}_n -
\int B_{n+1}^{\mathrm{H}} \drm\Phi_{\mathrm{rad}}$ and the latter
according to $B_{n+1}^{\mathrm{H}}$.
Note that the cancellation of singularities in these cross sections
requires a careful definition the \emph{additional PS subtractions}
$B_{n+1}^{\mathrm{S}}$; see \cite{Alwall:2014hca} for details.

Due to the high level of automation desired, compromises have been made when
generating the additional subtractions. These compromises have to be carried 
over when showering the a\textsc{MC@NLO} events, meaning that ME corrections 
cannot be used and that a different recoil scheme was adopted. The 
\emph{global recoil} scheme was designed for this purpose, in collaboration 
with the a\textsc{MC@NLO} authors. In this scheme, when one of the $n$ existing
partons branches, all the other $n -1$ partons share the recoil from 
giving the branching parton an off-shell mass. This is unlike the normal
\textsc{Pythia} final-state dipole shower, where only the colour-connected
partner takes a recoil. While the first emission has to be 
constructed with global recoil momentum sharing, everything beyond this
point could be done either way. Thus, \textsc{Pythia} allows to switch off 
global recoils at various stages in the evolution: 
\begin{enumerate}
\item after any emission has been 
produced with global recoil, 
\item after any emission of this parton has been produced
using global recoil, or 
\item after the parton multiplicity reaches a user-defined
limit. 
\end{enumerate}
In the first two options, no global recoil is applied for the hard 
events.

\subsubsection{Process-independent improvements}

\textsc{Pythia} also allows for process-independent improvements in order to
improve a wide range of observables simultaneously. The methods implemented
in or supported by \textsc{Pythia} fall into the category of multi-jet
merging schemes. These schemes use the parton shower when soft or collinear
partons are present, and fixed-order matrix elements for
well-separated partons. A consistent combination of all states with $n$
well-separated partons and $m$ soft-collinear partons is achieved for any
$m$ and $n$. States with any number of $n\leq N$ hard partons are described
with fixed-order accuracy. The definition of the boundary between 
soft/collinear and well-separated phase space regions requires a functional
form and a value for a cut called the \emph{merging scale}, 
$t_{\mathrm{MS}}$. Multi-jet merging methods in \textsc{Pythia} 
may be used to combine multiple LO or multiple NLO calculations.

The first native multi-jet merging scheme in \textsc{Pythia} is the CKKW--L 
method \cite{Lonnblad:2011xx}. Input LHEF samples for any
parton multiplicity and any process can be combined into a LO merged inclusive
sample. Overlaps between different multi-jet inputs are removed with the help
of Sudakov factors. These Sudakov factors are generated directly by the shower
after assigning a parton-shower history of on-shell intermediate states to 
the input state.
This history is obtained by reconstructing all possible ways in which the input
state could have been produced from the +zero-parton core process, and then
choosing amongst these evolution paths probabilistically. Together with the
generation of Sudakov factors directly from the shower and the inclusion of
$\alphas$ running and PDF rescaling, this ensures that no mismatch between
reweighted tree-level inputs and the parton shower is introduced, thus 
minimising the $t_{\mathrm{MS}}$ dependence. The inclusive cross section 
after CKKW--L merging then reads
\begin{equation}
B_{\mathrm{inc}} = \sum_{n=0}^{N-1} \widehat{B}_n + \widehat{B}_N ~.
\end{equation}
Here $\widehat{B}_n$ is the Born-level $n$-body cross sections, modified as
follows. First ME calculations are performed with a fixed $\alphas$
and PDF scale, and are here reweighted to apply to the scales of the 
reconstructed shower history. Secondly, Sudakov factors are introduced,
that remove the overlap between the tree-level cross sections and make them 
exclusive. The implementation on CKKW--L merging in \textsc{Pythia} supports 
arbitrary functional definitions of the merging scale. Three such different 
different choices are already included in the distribution. Other definitions 
can be introduced with the help of the \texttt{MergingHooks} class.

The mismatch between approximate virtual corrections introduced by 
parton-shower 
Sudakov factors and fixed-order radiation patterns leads to an unphysical 
dependence of inclusive cross sections on the merging scale in traditional
merging schemes. This issue can become severe for small merging scale values.
It is possible to correct the CKKW--L scheme to fix this problem
\cite{Lonnblad:2012ng,Platzer:2012bs}, leading to an add-and-subtract scheme 
to combine different multiplicities. Each
multiparton sample is reweighted as in CKKW--L, but instead of simply adding 
such samples, an all-order subtraction is included for each added sample.
The all-order subtractions can be generated on-the-fly, and in a 
process-independent fashion, with the help of PS histories:  
an $n$-parton state is processed a second time, the weight negated, 
and the shower is started from an $(n-1)$-parton state, which is
extracted from the PS history. The subtractions guarantee that all $n$-jet 
inclusive cross sections are separately independent of the merging scale. 
Also, the all-order subtractions complete the Sudakov factor of 
lower-multiplicity events. The resulting method, coined UMEPS 
\cite{Lonnblad:2012ng}, is also available natively in \textsc{Pythia}. 
UMEPS currently only supports one merging scale definition, namely the 
evolution scale $p_\perp$. Future upgrades of
the code could include additional merging scales, such that UMEPS can achieve
the same flexibility as CKKW--L.

A major step forward has been the introduction of NLO merging schemes 
extending the methods above to NLO accuracy for any number of jets
\cite{Lonnblad:2012ix,Frederix:2012ps,Hoeche:2012yf}. 
The basic construction principle of an NLO
merging scheme is to remove the approximate $\mathcal{O}(\alphas^{n+1})$ terms
from the $n$-parton samples of an LO merged calculation, and then add
back exclusive NLO calculations for all the samples for which terms have been
removed. Additional approximate NNLO corrections can then be introduced to
produce desirable effects, e.g.\ a stable prediction of inclusive cross 
sections.

The NLO extension of CKKW--L, called NL$^3$, is available natively in 
\textsc{Pythia} \cite{Lonnblad:2012ix}. Because of theoretical 
considerations, only the UMEPS merging scale definition is available. 
All fixed-order subtractions necessary to remove undesirable 
$\mathcal{O}(\alphas^{n+1})$ terms are performed on-the-fly. 
Subtractions of the numerical Sudakov factors are generated directly with the
parton shower, ensuring an implementation of phase-space limits and momentum
conservation that matches the parton shower exactly. As in CKKW--L, no 
approximate higher-order terms are adjusted.

The UNLOPS method, the NLO generalisation of UMEPS, is also available natively
in \textsc{Pythia} \cite{Lonnblad:2012ix}. Only the UMEPS merging scale 
definition is allowed. Also here, fixed-order subtractions are generated 
completely on-the-fly. To guarantee that all $n$-parton inclusive cross 
sections are given by
the $n$-parton NLO input calculations, improved approximate NNLO terms are 
introduced.

Both NL$^3$ and UNLOPS rely on inputs from NLO ME generators. These
inputs can be taken from POWHEG, MC@NLO or exclusive NLO calculations. For 
MC@NLO inputs, the first parton-shower emission has to be included already 
at the level of LHEF inputs, as would be the case for \textsc{Powheg-Box} 
inputs. Exclusive NLO inputs are available through the a\textsc{MC@NLO} 
package. Furthermore, auxiliary 
tree-level samples are necessary. It is possible to supplement even more 
tree-level samples for higher partonic multiplicities, making it possible
to combine NLO calculations up to a multiplicity $N$ with tree-level 
calculations for up to $M>N+1$ partons.

Another LO merging method introduced in \textsc{Pythia} is the
MLM prescription  \cite{Mangano:2006rw}. Here, the overlap of input samples 
with different partonic multiplicity is removed by jet counting and jet 
matching vetoes. 
Parton-shower-like $\alphas$-running effects are already included in the
input samples. \textsc{Pythia} supports MLM jet matching in the 
\textsc{Alpgen} scheme \cite{Mangano:2006rw,Cooper:2011gk}, 
the \textsc{Madgraph} scheme \cite{Alwall:2007fs}, and the
shower-$k_T$ scheme \cite{Alwall:2008qv}. The implementation of the 
jet-counting and jet-matching vetoes differs in each of these schemes.

MLM jet matching has been extended to NLO accuracy by the FxFx scheme
\cite{Frederix:2012ps}. This is available as a plugin to \textsc{Pythia}. 
FxFx combines multiple a\textsc{MC@NLO} calculations. Reweighting is 
necessary to remove the overlap of NLO calculations and include desirable 
higher-order corrections, e.g.\ $\alphas$-running. The 
$\mathcal{O}(\alphas^{n+1})$ subtractions, necessary to preserve the NLO 
accuracy, are done externally in the a\textsc{MC@NLO} program. A consistent 
interface 
to \textsc{Pythia} requires, as in the case for MC@NLO matching, the usage of 
the global recoil scheme. The jet counting and jet matching vetoes of 
the MLM prescription are amended to allow for an infrared safe definition. 
Only one matching criterion is available.

All native merging schemes in \textsc{Pythia} can be customised by the user
with the help of the \texttt{MergingHooks} structures: at crucial points in the
merging code, the user can access information and steer the code execution
directly. Please see section \ref{subsubsection:Matching-and-merging} and 
the online manual for more details.

\subsection{Other program components}

Standardised procedures have been introduced to link the program
to various external programs for specific tasks, see subsection 
\ref{subsec:external}.

Finally, some of the old jet finders and other analysis routines are
made available. Also included is a utility to generate, display and 
save simple one-dimensional histograms.

\subsection{Tunes}

The models for the various physics components of \textsc{Pythia} 
contain a number of parameters 
that have to be determined by comparisons with data. Such tunes have 
been produced both within the \textsc{Pythia} group and by the
experimental collaborations. Several of them are available by simple
master switches, so that not all parameters have to be set by hand.

The first tunes preceded LHC start-up and were mainly based on LEP and 
Tevatron data. However, due to uncertainties in the energy scaling,
they under-predicted the overall levels of MB and UE 
activity observed at the LHC by (10-20)\%. Later
tunes have included an increasing number of 
LHC measurements \cite{Katzy:2013lea}. Prior to \textsc{Pythia}~8.2,
the 4C tune \cite{Corke:2010yf} published in 2010 and 
including early 7~TeV LHC data, was the most commonly used
internally produced one, and it was the default in \textsc{Pythia}~8.1 
since version 8.145. It has been the starting point for several
subsequent tunes
by ATLAS \cite{ATLAS:2012xyz} and CMS \cite{CMS:2014xyz}.

The most recent tune that varies a larger number of parameters, and
that covers both LEP, Tevatron and LHC data, is the Monash 2013 one
\cite{Skands:2014pea}. It is the new default since
\textsc{Pythia}~8.200. 

Keep in mind that generators attempt to deliver a \emph{global}
description of the data; a tune is no good if it fits one distribution
perfectly, but not any others. For tuning purposes, 
it is therefore crucial to study the
simultaneous degree of agreement or disagreement over many, mutually
complementary, distributions. 
A useful online resource for making such comparisons can be found at
the \href{http://mcplots.cern.ch}{MCPLOTS} web
site~\cite{Karneyeu:2013aha}, which relies on the comprehensive
\textsc{Rivet} analysis toolkit~\cite{Buckley:2010ar}. The latter can also
be run stand-alone to make your own MC tests and comparisons.

Although \textsc{Pythia} may appear to have a
bewildering number of independently adjustable parameters, 
it is worth noting that most  of these only control
relatively small (exclusive) details of the event generation. The
majority of the 
(inclusive) physics is determined by only a few, very important ones, 
such as the effective values of $\alphas$, in the perturbative
domain, and  fragmentation-function and MPI
parameters, in the non-perturbative one.  

One would therefore normally take a highly factorised approach
to constraining the parameters, 
first constraining the perturbative ones, using IR-safe observables, 
and thereafter the non-perturbative ones, each ordered in a measure of
their relative significance to the overall modeling. This  allows one 
to concentrate on just a few parameters and a few carefully chosen 
distributions at a time, reducing the full parameter space to manageable-sized
chunks. Still, each step will often involve more than one single
parameter, and non-factorizable 
correlations may still necessitate additional iterations from
the beginning before a fully satisfactory set of parameters is
obtained.  

Recent years have seen the emergence
of automated tools that attempt to reduce the amount of both computer
and manpower required for this task, for instance 
by making full generator runs only for a
limited set of parameter points, and then interpolating between
these  to obtain approximations to what the true generator result
would have been for any intermediate parameter point, as, e.g., in 
\textsc{Professor}~\cite{Buckley:2009bj}. 
Automating the human expert input is more difficult. 
Currently, this is addressed by a combination of input solicited from
the generator authors 
and the elaborate construction of non-trivial weighting
functions that determine how much weight is assigned to each 
individual bin in each distribution. The field is still
burgeoning, and future sophistications are to be
expected. 
Nevertheless, at this point the overall quality of the tunes
obtained with automated methods appear to at least be competitive with
the manual ones. 

However, there are two important aspects
which have so far been neglected, and which it is becoming
increasingly urgent to address. The first is that an optimised tune is
not really worth much, unless you know what the uncertainty on the
parameters are. 
A few proposals for systematic tuning variations have
been made~\cite{Skands:2010ak,Richardson:2012bn}, but so far there is
no general comprehensive approach for 
establishing MC uncertainties by tune variations. The second issue is
that virtually all generator tuning is done at the ``pure'' LL shower
level,  and not much is known about what happens
to the tuning when matrix-element matching is subsequently
included. Due to the large processing power required, this issue 
is typically not accessible for individual users (or authors) to study,
but would require a dedicated effort with massive computing resources.   

Finally, rather than performing one global tune to all the
data, as is usually done, a more systematic check on the validity of
the underlying physics model could be obtained by instead performing
several independent 
optimisations of the model parameters for a range of different 
phase-space windows and/or collider environments.
In regions in which consistent parameter sets are obtained, e.g.\ with
reasonable $\Delta\chi^2$ values, 
the underlying 
model can be considered as interpolating well, i.e., it is universal. 
If not, a breakdown in the ability of the model to span different
physical regimes has been identified, and can be addressed, with the
nature of the deviations giving clues as to the nature of the breakdown. 
With the advent of automated tools,  
such systematic studies are now becoming feasible~\cite{Schulz:2011qy}. 

\section{Program Overview \label{sec:code}}

Also this section on code aspects is very brief, and only covers the 
main points, with emphasis on those that are new since 
\textsc{Pythia}~8.1. The 8.1 article \cite{Sjostrand:2007gs} offers 
a more extensive description, that in most respects is still valid. 

\subsection{Installation}

It is assumed that the code is to be installed on a Linux or Mac OS X system. 
After you download the \texttt{pythia8200.tgz} (or later) package 
from the \textsc{Pythia} web page,\\ 
\cindent \texttt{http://www.thep.lu.se/}$\sim$%
\texttt{torbjorn/Pythia.html}\\
you can unpack it with \texttt{tar xvfz pythia8200.tgz}, into a new 
directory \texttt{pythia8200}. The rest of the installation procedure 
is described in the \texttt{README} file in that directory. 
There is no explicit 
multiplatform support, but the self-contained character of the package 
should allow installation on any platform with a C++ compiler.

As first step, you should invoke the \texttt{configure} script, wherein 
notably you have to specify the location of all external libraries that 
you want to link \textsc{Pythia} to. You can also specify e.g.\ 
installation directories, whether you want to build a shared library 
in addition to the standard archive one, whether to turn off optimisation
to allow debugging, etc. If you do not plan to link to external libraries
and you accept the default choices this step can be skipped. Otherwise 
the \texttt{--help} argument provides a list of options.  Of particular 
interest are the linkages to \textsc{Fastjet}, HepMC, 
\textsc{LHAPDF}, and the gzip library for reading compressed LHE files.  
In the latter case, care must be taken to locate the boost library on 
a given platform.

As second step, a single \texttt{make} command invokes the 
\texttt{Makefile} to build and install libraries in a newly-created 
\texttt{lib} subdirectory, using information stored from the 
\texttt{configure} step in a \texttt{Makefile.inc} file.

There is also an optional third \texttt{make install} step, to make 
the program available to all local users, but this requires superuser 
privileges to execute. In this step the local \texttt{lib} contents 
by default are copied to \texttt{/usr/lib}, \texttt{include} to 
\texttt{/usr/include}, and \texttt{share} to \texttt{/usr/share}.
You can specify non-default directories in the \texttt{configure} step, 
e.g.\ to keep several versions accessible.

After this, the main program is up to the user to write. A worksheet
(found in the distribution) takes you through a step-by-step procedure, 
and sample main programs are provided in the \texttt{share/Pythia8/examples} 
subdirectory. These programs are included to serve as inspiration when 
starting to write your own program, by illustrating the principles involved. 
There is also a separate \texttt{Makefile} in the \texttt{examples} 
subdirectory, for linking the main programs to the \texttt{Pythia8} 
library and any other external libraries. 

The online manual is available if you open 
\texttt{share/Pythia8/htmldoc/Welcome.html} in your web browser. It will 
help you explore the program possibilities further. If you install the
\texttt{share/Pythia8/phpdoc} subdirectory under a web server you will 
also get extra help to build a file of commands to the \texttt{Settings} 
and \texttt{ParticleData} machineries, to steer the execution of
your main program.

\subsection{Program files and documentation}

The code in the \texttt{pythia8200} directory is subdivided into a set 
of files, mainly by physics task. Each file typically contains one main 
class, but often with a few related helper classes that are not used 
elsewhere in the program. Normally the files come in pairs: a \texttt{.h} 
header file in the \texttt{include/Pythia8} subdirectory and a 
\texttt{.cc} source code file in the \texttt{src} subdirectory. The new 
\texttt{include/Pythia8Plugins} subdirectory contains code pieces that
are not part of the core \textsc{Pythia} library but still can be of
general use, like nontrivial interfaces to other libraries.

The documentation is spread across four subdirectories to 
\texttt{share/Pythia8}: \texttt{xmldoc}, \texttt{htmldoc}, 
\texttt{phpdoc} and \texttt{pdfdoc}. Of these, the first is the most 
important one: the \texttt{xmldoc/*.xml} files contain all the settings 
and particle data, arranged by topic, and some further files contain 
e.g.\ PDF data grids. Therefore this directory must be accessible to 
the \texttt{Pythia} library. The program requires matching subversions 
of code and \texttt{xmldoc} files. For convenient reading in web browsers 
the \texttt{.xml} files are translated into a corresponding set of 
\texttt{.html} files in the \texttt{htmldoc} subdirectory and a set of 
\texttt{.php} files in \texttt{phpdoc}, to be accessed by opening the
respective \texttt{Welcome} file in a browser. The new \texttt{pdfdoc} 
directory collects the introductory text you are now reading, a 
worksheet/tutorial for beginners, and specialised descriptions of various 
physics aspects, the latter still at an early stage. 

A wide selection of main program examples are found in a fifth 
\texttt{share/Pythia8} subdirectory, \texttt{examples}.
Playing with these files is encouraged, to familiarise oneself with 
the program. For the rest, files should not be modified, at least not 
without careful consideration of consequences.
In particular, the \texttt{.xml} files are set read-only, and should 
not be tampered with, since they contain instructions from which 
settings and particle data databases are constructed. Any 
non--sensical changes here will cause difficult-to-track errors!

\subsection{Program flow}

The top-level \texttt{Pythia} class is responsible for the overall
administration, with the help of three further classes:
\begin{enumerate}
\item \texttt{ProcessLevel} is responsible for the generation of a 
process that decides the nature of the event. Only a very small 
set of partons/particles is defined at this level, so only the main 
aspects of the event structure are covered.
\item \texttt{PartonLevel} handles the generation of all subsequent 
activity on the partonic level, involving ISR, FSR, MPI and the 
structure of beam remnants.
\item \texttt{HadronLevel} deals with the hadronisation of this parton 
configuration, by string fragmentation, followed by the decays of 
unstable particles. It is only at the end of this step that realistic 
events are available, as they could be observed by a detector.
\end{enumerate}
At a level below these are further classes responsible for a multitude
of tasks, some tied to one specific level, others spanning across
them. 

Orthogonally to the subdivision above, there is another, more 
technical classification, whereby the user interaction with the
generator occurs in three phases.
\begin{itemize}
\item Initialisation, where the tasks to be performed are specified.
\item Generation of individual events, the event loop.
\item Finishing, where final statistics are made available.
\end{itemize}
Again the subdivision (and orthogonality) is not strict, with many 
utilities and tasks stretching across the borders, and with no 
finishing step required for many aspects. Nevertheless, as a rule, 
these three phases are represented by different methods  
inside the class of a specific physics task.

Information is flowing between the different program elements in
various ways, the most important being the event record, represented
by the \texttt{Event} class. Actually, there are two objects of this
class, one called \texttt{process}, that only covers the few partons
of the hard process of point 1 above i.e., containing information
corresponding to what might be termed the matrix element level,
and another called \texttt{event}, that covers the full story from the
incoming beams to the final hadrons. 

The \texttt{Settings} database keeps track of all integer, double,
Boolean and string variables that can be changed by the user to steer
the performance of \textsc{Pythia}, except that \texttt{ParticleData} 
is its own separate database. Various one-of-a-kind pieces of information
are transferred with the help of the \texttt{Info} class. 

In the following we will explore several of these elements further.

\subsection{The structure of a main program}

A run with \textsc{Pythia} must contain a certain number of commands. 
Notably the \texttt{Pythia} class is the main means of communication 
between the user and the event-generation process. We here present 
the key methods for the user to call, ordered by context. 

Firstly, at the top of the main program, the proper header file must
be included:\\
\cindent \texttt{\#include "Pythia8/Pythia.h"}\\
To simplify typing, it also makes sense to declare\\
\cindent \texttt{using namespace Pythia8;}\\ 
Given this, the first step in the main program is to create a 
generator object, e.g.\ with\\
\cindent \texttt{Pythia pythia;}\\
In the following we will assume that the \texttt{pythia} object
has been created with this name, but of course you are free to
pick another one.  

When this object is declared, the \texttt{Pythia} constructor initialises 
all the default values for the \texttt{Settings} and the
\texttt{ParticleData} databases. These data are now present in
memory and can be modified in a number of ways before the generator is
initialised. Most conveniently, \textsc{Pythia}'s settings and 
particle data can be changed by the two methods\\
\cindent \texttt{pythia.readString(string);}\\
for changing a single variable, and\\
\cindent \texttt{pythia.readFile(fileName);}\\
for changing a set of variables, one per line in the input file. 
The allowed form for a string/line will be explained as we consider 
the databases in subsection \ref{subsec:databases}. 

At this stage you can also optionally send in pointers to some external
classes, to hook up with user-written code or some external facilities, 
see subsection \ref{subsec:external}.

Once all the user requirements have been specified, a\\ 
\cindent \texttt{pythia.init();}\\
call will initialise all aspects of the subsequent generation.
Notably all the settings values are propagated to the various program 
elements, and used to precalculate quantities that will be used at later 
stages of the generation. Further settings changed after the 
\texttt{init()} call will be ignored, with very few exceptions. 
By contrast, the particle properties database is queried all the time, 
and so a later change would take effect immediately, for better or worse. 

Note, the \texttt{init()} method no longer accepts any
arguments. Rather, the user can set all initial conditions through
run-time parameters, and/or by explicitly settings pointers in the
user interface.

The bulk of the code is concerned with the event generation proper.
However, all the information on how this should be done has already
been specified. Therefore only a command\\
\cindent \texttt{pythia.next();}\\
is required to generate the next event. This method would be located
inside an event loop, where a required number of events are to be
generated. 

The key output of the \texttt{pythia.next()} command is the event
record found in \texttt{pythia.event}. A process-level summary of 
the event is stored in \texttt{pythia.process}. 

When problems are encountered, in \texttt{init()} or \texttt{next()}, 
they can be assigned one of three degrees of severity. 
\begin{itemize}
\item Abort is the highest. In that case the call could not
complete its tasks, and returns the value \texttt{false}. If this
happens in \texttt{init()} it is then not possible to generate any
events at all. If it happens in \texttt{next()} only the current event
must be skipped. In a few cases the abort may be predictable and
desirable, e.g.\ at the end-of-file of an LHEF. 
\item Errors are less severe, and the program can usually work around 
them, e.g.\ by backing up one step and trying again. Should that not 
succeed, an abort may result. 
\item Warnings are of informative character only, and do not require 
any corrective actions (except, in the longer term, to find more 
reliable algorithms).
\end{itemize}

At the end of the generation process, you can optionally call\\
\cindent \texttt{pythia.stat();}\\
to get some run statistics, both on cross sections for the 
subprocesses generated and on the number of aborts, errors and 
warnings issued.  

\subsection{The event record}

The \texttt{Event} class for event records is not much more than 
a wrapper for a vector of \texttt{Particle}s. This vector can expand 
to fit the event size. The index operator is overloaded, so that 
\texttt{event[i]} corresponds to the \texttt{i}'th particle of an 
\texttt{Event} object called \texttt{event}. For instance, given 
that the PDG identity code \cite{Beringer:1900zz} of a particle is 
provided by the \texttt{id()} method, \texttt{event[i].id()} returns 
the identity of the \texttt{i}'th particle. 

Line 0 is used to represent the event as a whole, with its total 
four-momentum and invariant mass, but does not form part of the
event history, and only contains redundant information. This line 
should therefore be dropped when you translate to another event-record 
format where the first particle is assigned index 1. It is only with 
lines 1 and 2, which contain the two incoming beams, that the history 
tracing begins. That way unassigned mother and daughter indices can be 
put 0 without ambiguity.

A \texttt{Particle} corresponds to one entry/slot/line in the event 
record. For each such particle a number of properties are stored,
namely
\begin{itemize}
\item the identity according to the PDG particle codes, 
\item the status (production reason, decayed or not),
\item two mother and two daughter indices (can represent ranges), 
\item a colour and an anticolour tag, 
\item the four-momentum and mass,
\item a production scale, 
\item a polarisation value, 
\item a Boolean whether a secondary vertex has been set, 
\item a four-vector representing the production vertex,
\item the invariant lifetime of the particle,
\item a pointer to the relevant particle data table entry, and 
\item a pointer back to the event the particle belongs to.  
\end{itemize}

From this information a multitude of derived quantities can easily be 
obtained; on kinematics, on properties of the particle species, on the 
event history, and more.  Note that particle status codes have changed
from the \textsc{Pythia}~6 and the record is organised in a different
way. In particular, \textsc{Pythia}~8 does not reprocess the
kinematics of the hard-process system after ISR; 
hence the original (Born-level) kinematic
configuration is preserved and can be retrieved from the record.

A listing of the whole event is obtained with \texttt{event.list()}. The 
basic identity, status, mother, daughter, colour, four-momentum and 
mass data are always given, but optional arguments can be set to provide 
further information, e.g.\ on the complete lists of mothers and daughters, 
and on production vertices.

The user would normally be concerned with the \texttt{Event} object that 
is a public member \texttt{event} of the \texttt{Pythia} class. Thus 
\texttt{pythia.event[i].id()} would be used to return the identity of 
the \texttt{i}'th particle, and \texttt{pythia.event.size()} to give 
the size of the event record. 

A \texttt{Pythia} object contains a second event record for the 
hard process alone, called \texttt{process}. This record is primarily
used as process-level input for the generation of the complete event.

The event record also contains a vector of junctions, i.e.\ vertices 
where three string pieces meet, and a few other pieces of information.

\subsection{Other event information}

A set of one-of-a-kind pieces of event information is stored in the
\texttt{info} object, an instance of the class
\texttt{Info}, in the \texttt{Pythia} class. This is mainly
intended for processes generated internally, but some of the information
is also available for external processes.

You can use \texttt{pythia.info} methods to extract information on, e.g.
\begin{itemize}
\item incoming beams,
\item the event type,
\item kinematics of the hard process,
\item values of parton distributions and couplings,
\item event weights and cross section statistics, 
\item MPI kinematics, and
\item some extra variables in the LHEF 3.0 standard.
\end{itemize}
The \texttt{info.list()} method prints information for 
the current event.

In other classes there are also methods that can be called to do a 
sphericity or thrust analysis, or search for jets with the $k_{\perp}$,
Cambridge/Aachen, anti-$k_{\perp}$ or other clustering algorithms
\cite{Salam:2009jx}. These take the event record as input.

\subsection{Databases \label{subsec:databases}}

Inevitably one wants to be able to modify the default behaviour of a 
generator. Currently there are two \textsc{Pythia}~8 databases with 
modifiable values. One deals with general settings, the other 
specifically with particle data. 

The key method to set a new value is\\
\cindent \texttt{pythia.readString(string);}\\
The typical form of a string is\\ 
\cindent \texttt{"variable = value"}\\
where the equal sign is optional and the variable begins with a letter 
for settings and a digit for particle data. A string not beginning with 
either is considered as a comment and ignored. Therefore inserting an 
initial !, \#, \$, \%, or another such character, is a good way to 
comment out a command. For non-commented strings, the match of the name 
to the database is case-insensitive. Strings that do begin with a letter 
or digit and still are not recognised cause a warning to be issued, unless 
a second argument \texttt{false} is used in the call. Any further text 
after the value is ignored, so the rest of the string can be used for 
any comments. For variables with an allowed range, values below the minimum 
or above the maximum are set at the respective border. For \texttt{bool} 
values, the notation \texttt{true} = \texttt{on} = \texttt{yes} = 
\texttt{ok} = 1 may be used interchangeably. Everything else gives 
\texttt{false}, including but not limited to \texttt{false}, 
\texttt{off}, \texttt{no} and \texttt{0}. 

The \texttt{readString(...)} method is convenient for changing one or two 
settings, but becomes cumbersome for more extensive modifications. In 
addition, a recompilation and relinking of the main program is 
necessary for any change of values. Alternatively, the changes can 
therefore be collected in a file where each line is a 
character string defined in the same manner as above without
quotation marks. 
The whole file can then be read and processed with a command\\
\cindent \texttt{pythia.readFile(fileName);}\\
As above, comments can be freely interspersed. Furthermore the 
\texttt{/*} and \texttt{*/} symbols at the beginning of lines can be 
used to comment out a whole range of lines.

\subsubsection{Settings}

We distinguish four kinds of user-modifiable variables, by the way
they have to be stored,
\begin{itemize}
\item a \texttt{Flag} is an on/off switch, and is stored as a 
\texttt{bool},
\item a \texttt{Mode} corresponds to an enumeration of 
separate options, and is stored as an \texttt{int},
\item a \texttt{Parm}, short for parameter, takes a 
continuum of values, and is stored as a \texttt{double},
\item a \texttt{Word} is a text string (with no embedded blanks)
and is stored as as a \texttt{string}. 
\end{itemize}
There are also the \texttt{FVec} vector of \texttt{bool}s, 
\texttt{MVec} vector of \texttt{int}s and  \texttt{PVec} vector of 
\texttt{double}s. Collectively the above kinds of variables are called
settings. Not surprisingly, the class that stores them is called 
\texttt{Settings}. 

Each variable stored in \texttt{Settings} is associated 
with a few pieces of information, typically
\begin{itemize}
\item the variable name, of the form \texttt{group:name} 
e.g.\ \texttt{TimeShower:pTmin},
\item the default value, set in the original declaration, 
\item the current value, and
\item an allowed range, represented by minimum and maximum values, 
where meaningful.
\end{itemize}
For the vector variants, default and current values are vectors, 
and have to be  manipulated as such, while the allowed range is stored
as scalars, i.e.\ shared by all the components.

Technically, the \texttt{Settings} class is implemented with the help 
of separate maps, one for each kind of variable, with the name
used as key. The default values are taken from the \texttt{.xml}
files in the \texttt{xmldoc} subdirectory at initialisation. The 
\texttt{settings} object is a public member of the \texttt{Pythia} class, 
and is initialised already in the \texttt{Pythia} constructor, such that 
default values are set up and can be changed before the \texttt{Pythia} 
initialisation. All public \texttt{Settings} methods can be accessed by 
\texttt{pythia.settings.command(argument)}. As already mentioned, for 
input the \texttt{pythia.readString(...)} method is to be preferred, 
since it also can handle particle data. A typical example would be\\
\cindent \texttt{pythia.readString("TimeShower:pTmin = 1.0");}\\
A vector can be read in as a comma-separated list.

You may obtain a listing of all variables in the database by calling\\
\cindent \texttt{pythia.settings.listAll();}\\
The listing is strictly alphabetical, which at least means that names
in the same area are kept together, but otherwise may not be so 
well-structured: important and unimportant ones will appear mixed.
A useful alternative is\\
\cindent \texttt{pythia.settings.listChanged();}\\
which will only print a list of those variables that differ from their 
defaults.

In user interfaces to \textsc{Pythia}, there may be cases when one
wants another method to set initial parameters. It can be cumbersome
and error-prone to translate various parameters into strings. In this
case, one can use the method
\texttt{pythia.settings.type(name,value)}, where \texttt{type} is
\texttt{flag}, \texttt{mode}, \texttt{parm}, or \texttt{word}, 
and \texttt{value} is a \texttt{bool}, \texttt{int},
\texttt{double}, or \texttt{string}, respectively. 

\subsubsection{Processes}

All internal processes available in \textsc{Pythia}~8 
can be switched on and off via the ordinary settings machinery, 
using flags of the generic type \texttt{ProcessGroup:ProcessName}. By
default all processes are off. A whole group can be turned on by a
\texttt{ProcessGroup:all = on} command, then overriding the individual
flags.

Note that processes in the \texttt{SoftQCD} group are of a kind 
that cannot be input via the LHA, while essentially all other kinds
could. 

Each process is assigned an integer code. This code is not used in
the internal administration of events; it is only intended to allow
a simpler user separation of different processes. Also the process 
name is available, as a string.

For many processes it makes sense to apply phase space cuts. 
Some simple ones are available as settings, whereas more sophisticated
can be handled with user hooks, see subsection \ref{subsubsec:userhooks}. 
In addition, for any resonance with a Breit-Wigner mass distribution, 
the allowed mass range of that particle species is taken into account, 
thereby providing a further cut possibility. Note that the 
\texttt{SoftQCD} processes do not use any cuts but generate their 
respective cross sections in full.   

\subsubsection{Particle data}

A number of properties are stored for each particle species:
\begin{itemize}
\item PDG identity code \cite{Beringer:1900zz},
\item name, and antiparticle name where relevant,
\item presence of antiparticle or not,
\item spin type,
\item electric charge,
\item colour charge,
\item nominal mass,
\item Breit-Wigner width,
\item lower and upper limits on allowed mass range,
\item nominal proper lifetime, 
\item constituent masses, specifically for quarks and diquarks,
\item scale-dependent running masses specifically for quarks,
\item whether a particle species may decay or not,
\item whether those decays should be handled by an external program,
\item whether a particle is visible in detectors (unlike neutrinos,)
\item whether it is a resonance with a perturbatively calculable width, 
and
\item whether the resonance width should forcibly be rescaled.
\end{itemize}
Methods can be used to get or set (most of) these properties.

Each particle kind  also has a a vector of decay channels associated 
with it. The following properties are stored for each decay channel:
\begin{itemize}
\item whether a channel is on or off, or on only for particles
or antiparticles,
\item the branching ratio,
\item the mode of processing this channel, possibly with 
matrix-element information, 
\item the number of decay products in a channel (at most 8), and
\item a list of the identity codes of these decay products.
\end{itemize}

Technically, the \texttt{ParticleData} class contains a map with the
PDG identity code used as key to the 
\texttt{ParticleDataEntry} storing the properties of a particle species. 
The default particle data and decay table is read in from the 
\texttt{xmldoc/ParticleData.xml} file at initialisation. The 
\texttt{particleData} object is a public member of the \texttt{Pythia} 
class. It is initialised already in the \texttt{Pythia} constructor, 
such that default values are set up and can be changed before the 
\texttt{Pythia} initialisation. All public \texttt{ParticleData} methods 
can be accessed by \texttt{pythia.particleData.command(argument)}.

As already mentioned, for input the \texttt{pythia.readString(string)} 
method is to be preferred, since it also can handle settings.
It is only the form of the \texttt{string} that needs to be specified
slightly differently than for settings, as\\ 
\cindent \texttt{id:property = value}\\ 
The \texttt{id} part is the standard PDG particle code, i.e.\ a number, 
and \texttt{property} is one of the ones already described above, 
with a few minor twists. In order to change the decay data, the 
decay channel number needs to be given right after the particle number, 
i.e.\ the command form becomes\\
\cindent  \texttt{id:channel:property = value}\\
As before, several commands can be stored as separate lines in a file, 
and then be read with \texttt{pythia.readFile(fileName)}.

For major changes of the properties of a particle, the above 
one-at-a-time changes can become rather cumbersome. Therefore 
a few extended input formats are available, where a whole
set of properties can be given after the equal sign, separated
by blanks and/or by commas. Notably (almost) all properties of a
particle or of a decay channel can be provided on a single line.

Often one may want to allow only a specific subset of decay channels
for a particle. This can be achieved by switching on or off channel
by channel, but a few smart commands exist that initiate a loop over 
all decay channels of a particle and allows a matching to be 
carried out. That way channels can be switched on/off for specific
inclusive or exclusive particle contents. There are also further methods 
to switch on channels selectively either for the particle or for the 
antiparticle.

When a particle is to be decayed, the branching ratios of the allowed 
channels are always rescaled to unit sum. There are also methods for 
by-hand rescaling of branching ratios.

You may obtain a listing of all the particle data by calling\\
\cindent \texttt{pythia.particleData.listAll()}.\\ 
The listing is by increasing \texttt{id} number. To list only those 
particles that have been changed, instead use\\ 
\cindent \texttt{pythia.particleData.listChanged()}.\\ 
To list only one specific particle \texttt{id}, use \texttt{list(id)}. 
It is also possible to \texttt{list} a \texttt{vector<int>} of 
\texttt{id}'s. 

\subsection{Links to external programs\label{subsec:external}}

While \textsc{Pythia}~8 itself is self-contained and can be
run without reference to any external library, often one does
want to make use of other programs that are specialised on some aspect
of the generation process. The HTML/PHP documentation accompanying the
code contains full information on how the different links should be set
up. Here the purpose is mainly to point out the possibilities that
exist.

For some of the possibilities to be described, \textsc{Pythia} contains 
a base class that the user can derive new code from. This derived class 
can then be linked by a command of the type \texttt{pythia.setXxxPtr(Yyy)}, 
where \texttt{Yyy} is a pointer to the derived class. The linking has 
to be performed before the \texttt{pythia.init()} call.

\subsubsection{The Les Houches interface}

The LHA \cite{Boos:2001cv}
is the standard way to input hard-process information from a 
matrix-elements-based generator into \textsc{Pythia}. The conventions 
for which information should be stored were originally defined in a 
Fortran context. To allow a language-independent representation, the 
LHEF was introduced \cite{Alwall:2006yp}. 
Subsequent to the original version 1.0, extended LHEF versions 2.0 
\cite{Butterworth:2010ym} and 3.0 \cite{Butterworth:2014efa} have 
been proposed. The current \textsc{Pythia} implementation is based on 
the 3.0 standard, which is backwards compatible with 1.0.

At the core of this implementation is the \texttt{LHAup} base class
which contains generic reading and printout methods, based on LHEF 1.0. 
It even allows for reading from gzipped LHE files. The derived 
\texttt{LHAupLHEF} class extends on this by handling LHEF 3.0, via a 
number of auxiliary program elements. Methods in the \texttt{Info} class
gives access to the new extra 3.0 information.
Of less interest, the derived \texttt{LHAupFromPYTHIA8} 
class allows you to write an LHEF with \textsc{Pythia}-generated 
hard-process events. 

You can create an \texttt{LHAup} object yourself and hand in a pointer 
to it. This can be used e.g.\ to provide a direct link to another program, 
such that one event at a time is generated and passed.

\subsubsection{Matching and merging}
\label{subsubsection:Matching-and-merging}

As mentioned in subsection \ref{subsec:matchmerge}, \textsc{Pythia} 
implements a wide variety of matching and merging procedures. Some of 
these form part of the core code, and only require simple LHEF input. 
Others require extra interfaces, e.g.\ for matching to \textsc{Powheg-Box} 
events, for input of \textsc{Alpgen} \cite{Mangano:2002ea} events that 
are not adhering to the LHEF standard, 
for MLM style multijet matching, and so on. This is a field in 
continued strong evolution, and so we refer to the online manual and
the example main programs for up-to-date details. For native
merging schemes (CKKW--L, UMEPS, NL$^3$ and UNLOPS), the possibility also
exists to write your own merging hooks class to tailor the 
merging procedure. This can be achieved by using the \texttt{MergingHooks}
structures, which give user access to the list below.

\begin{itemize}
\item The functional definition of the merging scale. This can be useful if
new merging criteria should be included. This option is currently only
supported by CKKW--L merging.
\item The weight associated to the event, allowing the user to change the
event weight or explicitly veto an event. This can be useful if the samples 
have been produced with severe cuts on the core process, that cannot be 
replicated in the high-multiplicity LHEF inputs.
\item The construction of all parton shower histories, by allowing the user
to disallow some reconstructed states. This can be useful when a certain class
of clusterings should be investigated.
\item The probability with which a history is picked, by allowing the user to
change the weight associated to the core scattering. This can be useful to
implement new hard matrix element weights into the merging.
\item The trial emissions which produce the Sudakov factors, by allowing the
user to ignore certain types of emissions. This can be useful to align
trial and regular showers if the regular shower has been constrained
to not produce certain emissions.
\end{itemize}

\subsubsection{SUSY parameter input}

\textsc{Pythia}~8 does not contain a machinery for calculating 
masses and couplings of SUSY particles from some small set
of input parameters. Instead the SUSY Les Houches Accord (SLHA)
\cite{Skands:2003cj,Allanach:2008qq} is used to provide this information, 
as calculated by some external program. You need to supply the name of 
the file where the SLHA information is stored, in an appropriate setting, 
and then the rest should be taken care of automatically. SLHA information
may also be embedded in the header block of an LHEF, and be read from 
there.

\subsubsection{Semi-internal processes and resonances}
\label{sec:semi-internal}

When you implement new processes via the LHA then all 
flavour, colour and phase-space selection is done externally, before your 
process-level events are input for further processing by \textsc{Pythia}. 
However, it is also possible to implement a new process in exactly the 
same way as the internal \textsc{Pythia} ones, thus making use of the 
internal phase-space selection machinery to sample an externally provided 
cross-section expression.

The matrix-element information has to be put in a new class that derives
from one of the existing classes for one-, two- or three-body final states.
Since \textsc{Pythia} does not have a good phase-space sampling machinery
for three or more particles, in practice we are restricted to  $2 \to 1$ 
and $2 \to 2$ processes. The produced particles may be resonances, 
however, so it is possible to end up with bigger final multiplicities 
through sequential decays, and to include further matrix-element weighting 
in those decays.

Once a derived class has been written, an instance of it can be handed in
to \textsc{Pythia}. From there on the process will be handled on equal 
footing with internally implemented processes. Interestingly, 
\textsc{MadGraph}~5 \cite{Alwall:2011uj} has the capability to generate 
complete such classes. 
 
If your new process introduces a new particle you have to add it and its
decay channels to the particle database, as already explained. To obtain 
a dynamical calculation, however, where the width and the branching ratios 
can vary as a function of the currently chosen mass, you must also create 
a new class for it that derives from the \texttt{ResonanceWidths} class 
and hand in an instance of it.

\subsubsection{Parton distribution functions}

In addition to the built-in PDFs, a larger selection can be obtained via 
the interfaces to the \textsc{LHAPDF5} 
\cite{Whalley:2005nh} or \textsc{LHAPDF6} 
\cite{Buckley:2014xyz} library. Should this not be enough, it is 
possible to write your own classes derived from the \texttt{PDF} base class, 
and hand them in. You can hand in one PDF instance for each incoming beam, 
and additionally have separate PDFs for the hard process and for Pomerons, 
i.e.\ altogether up to six different input PDFs.

\subsubsection{Parton showers}
 
It is possible to replace the existing timelike and/or spacelike showers
in the program by your own. This is truly for experts, since it requires
a rather strict adherence to a wide set of rules. These are described
in detail in the HTML/PHP documentation. The \textsc{Vincia} program 
\cite{Giele:2013ema} offers a first example of a plug-in of an external 
(timelike) shower. 

\subsubsection{Decay packages}

While \textsc{Pythia} is set up to handle any particle decays,
decay products are often (but not always) distributed isotropically 
in phase space, i.e.\ polarisation effects and nontrivial matrix
elements usually are neglected in \textsc{Pythia}. Especially for the 
$\mathrm{B}$ mesons it is therefore common practice to rely on 
the dedicated \textsc{EvtGen} decay package \cite{ Ryd:2005zz}. In the 
past also \textsc{Tauola} was used for tau lepton decays 
\cite{Jadach:1993hs}, but now \textsc{Pythia} contain its own detailed 
tau decay handling, so that is less of an issue. 

The \texttt{DecayHandler} is a base class for the external handling of 
some decays. The \texttt{decay(...)} method in it should do the decay
for a specified list of particles, or return \texttt{false} if it fails. 
In the latter case \texttt{Pythia} will try to do the decay itself. Thus 
one may implement some decay channels externally and leave the rest for 
\texttt{Pythia}, assuming the \texttt{Pythia} decay tables are adjusted 
accordingly.  

The \textsc{Photos} program \cite{Davidson:2010ew} is often used to add QED
radiative corrections to decays. Currently there is no dedicated interface
for this task. Instead such corrections can be imposed on the 
event record, after \textsc{Pythia} has completed the task of fully
generating an event. The reason this works is that the emission of
a photon does not change the nature of the other particles in a
decay, but only slightly shift their momenta to compensate. 

\subsubsection{Beam shape}

It is possible to write your own \texttt{BeamShape} class to select the 
beam momentum and the interaction vertex position and time event-by-event. 
The default is to have no momentum spread and put the primary vertex at 
the origin, while the preprogrammed alternatives only give simple Gaussian 
approximations for the spread of these quantities.

\subsubsection{Random-number generators}

\texttt{RndmEngine} is a base class for the external handling of 
random-number generation. The user-written derived class is called 
if a pointer to it has been handed in. Since the default 
Marsaglia-Zaman algorithm \cite{Marsaglia:1990xyz} is quite good, 
there is no physics reason to replace it, but this may still be required 
for consistency with other program elements in big experimental frameworks.

\subsubsection{User hooks \label{subsubsec:userhooks}}

Sometimes it may be convenient to step in during the generation process: 
to modify the built-in cross sections, to veto undesirable events or 
simply to collect statistics at various stages of the evolution. There is 
a base class \texttt{UserHooks} that gives you this access at some
selected places of the code execution. This class in itself does nothing; 
the idea is that you should write your own derived class for your task. 
A few very simple derived classes come with the program, mainly as 
illustration.

The list of possibilities is slowly expanding with time, and currently
includes eight sets of methods that can be overloaded.
\begin{itemize}
\item Ones that gives you access to the event record in between the 
process-level and parton-level steps, or in between the parton-level 
and hadron-level ones. You can study the event record and decide whether 
to veto this event.
\item Ones that allow you to set a scale at which the combined MPI, 
ISR and FSR downwards evolution in $\pT$ is temporarily interrupted, 
so the event can be studied and either vetoed or allowed to continue the 
evolution.
\item Ones that allow you to to study the event after the first few 
ISR/FSR emissions, or first few MPI, so the event can be vetoed or 
allowed to continue the evolution.
\item Ones that allow you to study the latest initial- or final-state 
emission and veto that emission, without vetoing the event as a whole.
\item Ones that give you access to the properties of the trial hard 
process, so that you can modify the internal \textsc{Pythia} cross section, or
alternatively the phase space sampling, by your own correction factors.
\item Ones that allow you to reject the decay sequence of resonances 
at the process level.
\item Ones that let you set the scale of shower evolution, specifically 
for matching in resonance decays.
\item Ones that allow colour reconnection, notably in the context of 
resonance decays. 
\end{itemize}

\subsubsection{Jet Finders}

The \textsc{Pythia} package contains a few methods historically used to
characterise $\erm^+\erm^-$ annihilation events, including some jet 
finders. The \texttt{SlowJet} class implements the 
$\prm\prm$-physics-oriented $k_{\perp}$, Cambridge/Aachen and
anti-$k_{\perp}$ clustering algorithms \cite{Salam:2009jx}. 
The native implementation is slower than the \textsc{FastJet} one 
\cite{Cacciari:2011ma}, but the default now is to use \texttt{SlowJet} 
as a frontend for the FJcore part of \textsc{FastJet} package. The FJcore code 
is distributed together with the \textsc{Pythia} code by permission 
from the authors. There is also an interface that inputs \textsc{Pythia} 
events into the full \textsc{FastJet} library, for access to a wider set of 
methods, but then \textsc{FastJet} must be linked.    

\subsubsection{The HepMC event format \label{subsubsec:hepmc}}

The HepMC \cite{Dobbs:2001ck} event format is a standard format
for the storage of events in several major experiments. The translation
from the \textsc{Pythia}~8 \texttt{Event} format should be done
after \texttt{pythia.next()} has generated an event. Therefore there
is no need for a tight linkage, but only to call the relevant
conversion routine from the main program written by the user.
Currently HepMC version 2 is supported, and a separate interface
to version 3 \cite{Butterworth:2014efa} is foreseen once this new
standard has reached a stable form.

\section{Outlook \label{sec:outlook}}

While the \textsc{Pythia}~8.100 release involved brand new code, 
with some relevant components still not fully in place, much has
happened since, and so the 8.200 one is of a much more mature and 
tried code. For applications at hadron colliders and for 
$\erm^+\erm^-$ annihilation there is no reason to cling on 
to \textsc{Pythia}~6.4, since 8.2 offers a complete replacement, 
with several improvements. The areas where 6.4 may still be useful
are  $\erm\prm$, $\gamma\prm$ and $\gamma\gamma$,
which still are lacking in 8.2. They will be added when time permits,
but have lower priority than the exploration of LHC data, and 
improvements that may spring from new physics needs or insights here. 
In addition the code will have to evolve to match other high-energy 
physics libraries. The program will therefore continue to be 
developed and maintained over the years to come. 

\section*{Acknowledgements}

Work supported in part by the Swedish Research Council,
contract number 621-2013-4287, and in part by the MCnetITN
FP7 Marie Curie Initial Training Network, contract
PITN-GA-2012-315877.\\
The help of numerous users is gratefully acknowledged,
in terms of code contributions, bug fixes and helpful comments;
their significant impact can be gleaned from the Update History
of the \textsc{Pythia}~8.1 distribution.

\end{document}